# Bessel beams, propagationally invariant beams, and axicons: An historical and tutorial review

Colin J. R. Sheppard
*Honorary Professor*
*School of Optometry and Vision Science, UNSW Sydney, NSW 2052, Australia*
e-mail: colinjrsheppard@gmail.com

The Bessel beam is discussed, including its historical development. We present the relationship with the Arago Spot, annular pupils, axicons and pupil filters. We also discuss Bessel-Gauss beams; higher order Bessel beams; electromagnetic polarization properties; waveguide and resonator modes; polychromatic and pulsed beams; one-dimensional Airy beams; extended depth of focus; and light sheets and photonic nanojets.

## 1 Introduction

This paper is dedicated to Professor Kehar Singh on his 84[th] birthday. In the 1970s and 1980s, it was common practice to send postcards requesting reprints of journal papers. I received several such requests from Kehar Singh for reprints of my work on Bessel beams and confocal imaging, and as a consequence visited him in 1981. I learnt much optics from studying his impressive collection of reprints. I also visited on that trip Professor De at Calcutta University, and Professor Sirohi at IIT Madras.

The topic of Bessel beams, and other propagationally invariant beams, is of great and growing interest, due partly to their strange behaviour, and partly to their widespread uses, in imaging, optical metrology, communications, trapping and manipulation, material processing and lithography, waveguides and resonators, and so on. The Author's interest in Bessel beams started in 1974, when he started researching in laser scanning microscopy at Oxford University. A paper 'Imaging properties of annular lenses' was published in 1979 [1]. The original submitted article had a comprehensive review of publications up to that time, but was removed on the advice of a reviewer. Most recent papers fail to cite the early work, so the aim in this review is to provide a treatment of the early research. We concentrate on papers discussing the beams themselves rather than their applications.

In 1977 we published two papers on lenses with annular aperture (one with experimental results) [2, 3], and in 1978, we published a further two papers that clearly describe the propagational invariance of a Bessel beam [4, 5]. The first 1978 paper, which was mainly concerned with the electromagnetic (polarization) properties of Bessel beams (see Section 10), says [4]:

'The intensity distribution near the focus of a lens system with annular aperture has been studied by Linfoot and Wolf [6]. It is found that the variation along the optic [*sic*] axis is 'stretched', until in the limit as the annulus becomes very narrow, the intensity becomes constant in the direction parallel to the optic axis (in the region in which the calculations [in the Debye approximation] are valid), thereby greatly increasing the depth of focus. The intensity distribution in the focal plane also approaches a limit as the annulus becomes narrow.'

The second paper 1978 paper introduced the Bessel-Gauss beam (see Section 8), and says [5]:

'The radial distribution for a $\delta$-ring is given by a zero-order Bessel function in any plane … perpendicular to the optic axis. That this is so is not surprising because such a wave is the circularly symmetric mode of free space. We are acquainted with modes of this form in circular waveguides, and we can consider free space as the limiting case of a waveguide of very large diameter. Such an overmoded waveguide has an infinity [meaning a continuum] of circularly symmetric modes, that is the scale of the Bessel functions may be chosen at will. A wave with zero-order Bessel-function radial distribution propagates without change.'

A patent was filed on the use of Bessel beams in scanning microscopes [7].

Bessel beams are not physically realizable, as they contain infinite energy. This is not as much of a problem as it may seem, as even plane waves are also not physically realizable but often used as an idealization.

There are a few fundamentally different ways of generating approximations to Bessel beams, as illustrated in Fig. 1. In (a), a narrow annular pupil (annular slit) is shown in the front focal plane of a lens [8, 9]. The rays in a meridional plane after focusing by the lens are all parallel, thus forming a Bessel beam, with cross sectional intensity $J_0^2$. Of course, this is an idealization, as if the annulus were infinitely narrow, the lens aperture would have to be infinitely large to accept all the rays from the annulus. The rays are equivalent to plane waves, so the Bessel beam is formed by an angular spectrum of plane waves incident along the surface of a cone. As the plane of observation moves along the optical axis, the phase of all the plane wave components all change by the same amount, so the cross section of the beam is invariant, and all that changes is the absolute phase. In part (b) of the figure, an axicon (conical prism), as proposed by McLeod, is illuminated with a plane wave [10, 11]. Again, rays are deflected so that they are parallel to each other, at an angle $\alpha$ to the optical axis. The axicon is shown with its conical surface closest to the source, whereas most papers on axicons show the plane surface to be nearest to the source. We would expect the arrangement shown to give reduced chromatic aberration, in a similar way to the minimum deviation alignment of a prism. However, as we discuss later in Section 4, the diffraction behaviour of the axicon depends critically on the effect of the vertex region, near the optical axis. Note that an axicon is situated in the Fresnel region, whereas an annulus is in the Fourier plane. In part (c) of the figure, a grating (diffractive axicon) is used, instead of a refractive axicon [12]. The grating is analogous to a zone plate, but as the rays after diffraction are parallel, the spacing of the grating is linear, rather than quadratically decreasing as in a conventional zone plate. The grating could be a phase grating, and as it is in the Fresnel regime can be regarded as a hologram [13, 14]. Interestingly, Dyson's paper predates McLeod's. Part (d) of the figure shows a more general binary filter in the front focal plane of a lens [15, 16]. This filter can be regarded as a Fourier transform hologram.

We continue by discussing in Section 2 the well-known Arago spot, and then to consider these different approaches, in Sections 3, 4, and 9. We also discuss higher order Bessel beams (Section 7); Bessel-Gauss beams (Section 8); polarization (electromagnetic) properties (Section 10); waveguide and resonator modes (Section 11); polychromatic and pulsed beams (Section 12); one-dimensional (1D) beams, Airy beams, extended depth of focus, and light sheets (Section 14): and photonic nanojets (Section 16). In each of these sections, the citations are arranged mostly in historical order.

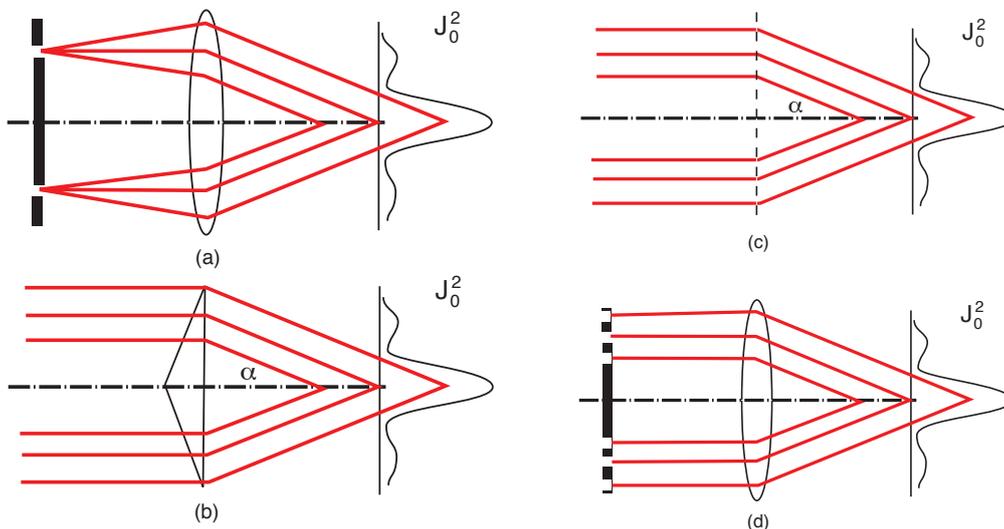

Fig. 1. Ways of generating approximations to a Bessel beam. (a): An annular pupil in the front focal plane of a lens. (b): An axicon (conical prism).(c): A Bessel zone plate, equivalent to a holographic filter. (d): A binary filter in the front focal plane of a lens.

## 2 Arago spot

A bright spot with a $J_0$ Bessel field-strength cross section is observed on the optical axis, in the shadow region, behind a disk shaped obstacle. This is known as the Arago, or Poisson, spot. The spot, which is an approximation to a Bessel beam, is of constant peak intensity along the axis in the Fresnel diffraction region.

The Arago spot was reported experimentally by Delisle in 1715, and by Maraldi in 1723, well before the wave theory of light was well established. Then Arago also observed it (in 1818), and this was taken as experimental evidence for the wave theory of light. The Arago spot is an approximation to a Bessel beam. So, far from being a 'non-diffracting' or a 'diffraction-free beam', the Bessel beam is actually *caused* by diffraction. The propagationally invariant nature of the Bessel beam is a case of dynamic equilibrium between outward diffraction, and inward diffraction from the strong ring structure.

Also, the reason that the Arago spot was considered evidence for the validity of the wave theory was that the spot is in the geometrical shadow region, and caused by light that bends around the obstacle by diffraction. Light does not travel along the optical axis, but is replenished by light scattered from the boundary of the obstacle. So the concept of a Bessel beam 'reconstructing' or 'self-healing' after an obstruction has been known for 200 years.

Potter observed the ring structure of the Arago spot in 1841[17]:

'When the whole was adjusted, on looking through an eyelens of about 1 inch focal length, at the centre of the shadow cast by the disk, there was seen a bright central spot, of a white color slightly tinged with brown, surrounded by a greater or lesser number of coloured rings, according to the size of the disc.'

Lommel derived expressions for the amplitude in the paraxial Fresnel diffraction pattern of a circular aperture and a circular obstacle, illuminated with a general spherical wave, in terms of what are now known as Lommel functions [18]. For the circular obstacle, he states:

'...the light intensity at every point on the axis of the geometrical shadow is the same as the intensity of the unobstructed wave as if the obstacle were not present.'

Near the axis, the Lommel function $V_0$ tends to $J_0$, so, for illumination of the obstacle with a plane wave, the amplitude at the point $\rho, z$ in cylindrical coordinates, is

$$U(\rho,z) = \exp(ikz)\exp\left(\frac{ika^2}{2z}\right) J_0\left(\frac{ka\rho}{z}\right), \tag{1}$$

where $k = 2\pi/\lambda$ and $a$ is the radius of the obstacle. So, although the intensity along the axis is constant, the width of the Bessel function increases linearly with distance.

If the more accurate first Rayleigh-Sommerfeld diffraction formula is used, we obtain for the amplitude near the axis,

$$U(\rho,z) = \frac{z}{\sqrt{z^2+a^2}} \exp\left(ik\sqrt{z^2+a^2}\right) J_0\left(\frac{ka\rho}{z}\right), \tag{2}$$

so that the intensity decreases near to the obstacle, approaching zero at the obstacle. The form of the equation for the amplitude in this case agrees with an integration of the wave scattered by the edge of the disk, as predicted by the boundary-diffraction wave theory.

Porter produced an image of a triangular source using an obstacle [19]:

'It does not. seem to have been recognized that, as ordinarily exhibited, the bright spot is in reality an image of the source of light. … We have thus the possibility of obtaining sharp images of small objects, *using an opaque disk instead of a lens* [original author's italics].'

Rankine showed that the image produced by a disk-shaped obstacle exhibits aberrations, and proposed that these can be avoided by using a spherical obstacle because, by symmetry, a sphere gives no aberrations [20].

Raman and Krishnan compared diffraction by a disk and a sphere. They showed that, close to the obstacle, a disk produces a brighter spot than a sphere [21]:

'The disturbance incident on the surface of the sphere has to creep round it, as it were, over the arc XY before the rays diffracted out by the sphere can reach the point of observation, and must suffer a very considerable diminution in the process.'

Berko *et al.* demonstrated the Arago spot using laser illumination [22].

Harvey and Forgham analyzed the field behind a circular obstacle illuminated by a plane wave or a Gaussian beam, using Rayleigh-Sommerfeld diffraction theory. They compared the intensity along the axis with that for a circular aperture and an annular aperture. They also proposed that observation of the Arago spot could be used to investigate aberrations, especially astigmatism, of a laser beam from an annular resonator [23].

Harrison *et al.* described using a positive lens to image the Arago spot to a finite distance [24]. In this way, if the distance of the lens from the obstacle is equal to the focal length of the lens, the width of the Bessel function $J_0$ is independent axial position. This geometry is then similar to a lens with an annular pupil, giving a Bessel beam along the axis in the shadow region of the obstacle. Similarly, if the obstacle is illuminated with a convergent wave, the Arago spot is a Bessel beam, within the region that the Fresnel diffraction theory is valid. For illumination by a collimated wave, the Bessel beam is formed at infinity.

Saari *et al.* observed experimentally the formation of the Arago spot behind a circular obstacle for pulsed illumination in the time domain, and compared the results with theoretical predictions of the boundary-diffraction wave theory [25]. They found that the Arago spot, caused by scattering from the edge of the obstacle, propagates along the axis behind the wave front of the transmitted geometrical wave, but catches up with the geometrical wave at infinity because its velocity is superluminal. They compared the result with that for a circular aperture, where an Arago spot is still formed, and is temporally resolved from the geometrical wave on the axis.

The boundary waves interfere with each other and with the directly transmitted pulse, but the interference maximum on the axis (actually a temporally resolved spot of Arago) lags behind the direct pulse, and eventually catches up with it.

**3 Annular pupils**

The intensity in the focal plane of a lens was considered by Airy in 1835 [26]. He calculated by a power series expansion using Cartesian coordinates what is now known as the Airy disk. Although the Airy disk is now usually expressed in terms of a Bessel function $J_1$, his paper predates the widespread use of that name. Even though Bessel's paper was published in 1824 it gave numerical values of $J_0, J_1$ only up to an argument of 3.2. Airy goes on to calculate the effect of a circular obscuration, and the radii of the dark rings for a narrow annulus. He observes that

'Thus the magnitude of the central spot is diminished, and the brightness of the rings increased, by covering the central parts of the object-glass.'

Then in a second paper, in 1841, the same year as that of Potter, he used cylindrical coordinates to calculate the amplitude in the diffraction pattern of an elemental annular aperture (given by what is now known as the Bessel function $J_0$), giving a table of values, and goes on to integrate over the two cases of a circular aperture and an opaque disk [8]. Although Airy gave the expression for diffraction by an arbitrary annular aperture of finite cross section, he did not give any numerical results or plots for the general case.

Lord Rayleigh proposed that an annular pupil would allow the focal intensity to be reduced for observation of bright objects such as the sun, while maintaining the resolving power and also reducing the spherical aberration [9]. He stated explicitly that the intensity for a narrow annular pupil (`marginal rim') is given by $J_0^2$. He compared the intensity in the focal plane of a lens with a circular or a narrow annular pupil, and noted that the annular case gives a smaller focal spot, but with stronger outer rings. He noted that the maxima of $J_0^2$ occur at the zeroes of the Airy disk. In a post script added in *Scientific Papers*, he mentioned Airy's 1841 paper, `with which [he] was previously unacquainted', and went on to explain:

'There is no pretence of originality in the mathematics; indeed, the solution of the problem for the annulus is a [usual] step in the treatment of the question of a lens with full aperture, the details of which are well known.'

In 1888 he said [27]:

'It has been found by Herschel and others that the definition of a telescope is often improved by stopping off part of the central area of the object-glass; but the advantage to be obtained in this way is in no case great, and anything like a reduction of the aperture to a narrow annulus is attended by a development of the external luminous rings sufficient to outweigh any improvement due to the diminished diameter of the central area.'

and

'It may be instructive to contrast this with the case of an infinitely narrow annular aperture where the brightness is proportional to $J_0^2(z)$. When $z$ is great, ... the mean brightness varies as $z^{-1}$; and the integral $\int_0^\infty J_0^2(z) z \, dz$ is not convergent.'

So it seems Rayleigh knew that a 'Bessel beam' is non-physical.

Steward considered the effect of small aberrations on focusing using lenses with circular or annular pupils, and gives some plots of the intensity in the focal region [28, 29]. He gave a plot of the $J_0^2$ intensity in the focal plane for a narrow annular pupil [28], reproduced in Born and Wolf [30]. He showed that the aberrations are mostly reduced by using an annular aperture, and in particular that the depth of focus can be increased without limit [28]:

'From this formula it is evident that the distance between successive dark points upon the axis is given by ... $1/(1-\varepsilon^2)$; this tends to infinity as $\varepsilon$ tends to unity, as, indeed, is obvious from the consideration that in the limiting case the aperture reduces to a circular rim.'

Here we have changed the symbol for fractional radial obscuration to be $\varepsilon$. Although most aberrations are reduced for the annular case, he commented that

'The annular aperture is seen to be unfavourable in the case of astigmatism - at least as far as the central intensity is considered.'

He also considered semicircular pupils.

Dunham considered the resolution of a microscope with central stops [31].

Linfoot and Wolf presented contour plots of the intensity in the focal region for an annular pupil [6]. It should be noted that their calculation is based on the paraxial Debye approximation, i.e. for a spherical wave converging on to an aperture, so that their calculations are valid 'only when $u/4\pi$ does not become large compared with unity'. ($u$ is the axial optical coordinate, $u = 8\pi z n \sin^2(\alpha/2)/\lambda$, where $n \sin \alpha$ is the numerical aperture.). At this time the effects of Fresnel number on focusing were not well appreciated, even though the optical design community knew very well that placing the pupil in the front focal plane (rather than in the converging wave) ensures that the Debye theory is valid for all space behind the lens.

Steel investigated the effects of small aberrations on imaging with annular pupils [32]. He investigated different objects including disks, lines and edges, and calculated the optical transfer function (OTF). He showed that the effect of all aberrations disappears for a narrow annulus, except that for astigmatism it is increased by a factor $(1+\varepsilon^2+\varepsilon^4)^{1/2}$, i.e. by $\sqrt{3}$ for a narrow annulus. He discussed application to mirror microscopes.

O'Neill calculated analytically the OTF by integration of the intensity point spread function (PSF), and gave plots [33]. He concluded: 'Finally, for a thin annular ring, one might predict from the shape of the curves that as $\varepsilon$ approached unity, a sharp peak near the resolution limit of the system could be obtained.'

Riesenberg discussed the mirror microscope, using an annular objective [34].

Taylor and Thompson measured experimentally the intensity along the axis for an annular pupil [35]. 'The annulus was constructed by placing the aperture and the stop on a piece of uniformly thin mica.' The pupil was placed in close proximity to the lens [i.e. not in the front focal plane]. They demonstrated that the axial minima in intensity became more separated as the obscuration was increased.

Welford studied theoretically the use of annular apertures to increase focal depth [36]. He concluded that 'the increase in brightness in the outer diffraction rings would result in great loss of contrast in the images of extended objects.'

Asakura and Barakat investigated the effect of combined spherical aberration and defocus on focusing with an annular aperture [37]. They also considered an annular obscuration, as in a dark field microscope.

Kelly was perhaps the first to recognize the equivalence of an annular pupil and an axicon [38]. Unfortunately we have only his abstract, for a presentation at the OSA Annual Meeting:

'In polar coordinates the analog of a double slit is a very thin annulus, and the correpording Fraunhofer diffraction pattern is the square of a Bessel function of the first kind, zero order. The same pattern can be obtained without an annulus, and therefore at a considerably higher brightness level, by means of an axicon (i.e., the circular analog of the Fresnel biprism). The spatial frequency components and "harmonic purity" of this pattern are considered in terms of the Fourier transform of the squared Bessel function in polar coordinates.'

Mahan et al. and Izatt calculated the diffraction patterns of apertures bounded by arcs and radii of concentric circles [39, 40].

Barakat and Houston calculated the OTF for annular apertures with spherical aberration and defocus [41].

Mehta calculated the total illumination with radius (encircled energy) in the PSF for annular apertures as a series solution [42]. Tschunko also calculated (numerically) the encircled energy [43, 44]. He also showed that the outer rings are modulated in groups of $N_1, N_2$: for $\varepsilon \leq 1/3, N_2 = 1/\varepsilon$, $\varepsilon \geq 1/3, N_1 = 2/(1-\varepsilon)$ [45]. He explained this phenomenon by the interference of two waves scattered by the inside and outside of the annulus. Wiener also gave a plot of encircled energy for annular apertures [46] Goldberg and McCulloch calculated the energy contained in the few brightest rings, and also the encircled energy for illumination by a finite-sized incoherent circular source [47]. They pointed out that there is a reciprocal relationship between source and detection radius, and Otterman considered the special case of the energy within the geometrical image of the source [48]. Later, Stamnes et al. calculated analytically the encircled energy for annular apertures [49].

Sodha et al. considered the fluctuations in intensity for focusing with an annular aperture through a turbulent medium [50].

Singh and Kavathekar calculated images of periodic bar patterns with an annular pupil [51]. Dhillon and Singh investigated diffraction of partially coherent light by an annulus [52].

McCrickerd calculated, analytically by autocorrelation of the pupil function, the OTF for a narrow (e.g. $\varepsilon = 0.96$) annulus, and gave an approximate expression for the OTF for large obscuration [53]. He described that the fringes in the PSF are averaged out by a spread of wavelenths. He showed that the OTF is simply attenuated by defocusing. He proposed combining an annular pupil with coherent optical processing, and demonstrated equalization of the OTF.

Sirohi and Bhatnagar, De and Basway, and McKechnie calculated the two-point resolution for a microscope with an annular condenser [54-56]. Nayyar and Verma calculated the two-point resolution for a microscope with annular condenser and annular objective pupils [57]. Rose proposed using annular bright field imaging in the scanning transmission electron microscope [58].

Fenneman and Cruise calculated using an integral equation the optimum shape of an annular pupil to maximize the energy within a circular target [59]. They showed the equivalence of the problem to that of the hole-coupled resonator (see Section 12). They considered annuli with $\varepsilon$ up to 0.999. They concluded:

'The far-field pattern of an annulus can be viewed as both a diffraction and an interference phenomenon (somewhat like the double slit). The interference effect, which is reduced by optimum pupil shaping, becomes more pronounced as $\varepsilon$ increases. For example, for $\varepsilon = 0.8$ it takes [10 rings] to capture 90% of the light from a uniformly illuminated aperture. Optimum shaping puts 98% on the same size. Because the optimum pupil functions maximize the light on the target, they minimize the light outside.'

Nayyar and Verma calculated the diffraction pattern of an annulus illuminated with a Gaussian beam [60].

Sheppard investigated imaging in a scanning microscope with a narrow annular illumination pupil, observing [2].

'Close examination of the results of Linfoot and Wolf shows that in this way the (depth of focus is improved over and above the factor of $1/(1-\varepsilon^2)$, as the "wings" of the pattern for the full aperture are absent in that for the annular aperture.'

An experimental image of an edge object using an objective with a narrow annular pupil was given, showing that the contrast was very poor [2]. The use of a narrow annulus in the illumination arm of a confocal microscope was proposed. The confocal microscope exhibits a PSF given by the product of the illumination and detection PSFs, and as the maxima of the Bessel beam, $J_0$, coincide with the zeroes of $J_1$ in the Airy disk (this was pointed out by Rayleigh [9]), the outer rings are very weak, while the central lobe is made narrower.

Sheppard and Choudhury gave the OTF for a Bessel beam [3]. They gave a theoretical treatment of image formation in scanning microscopes [3]. Scanning microscopes with a large area detector behave in a similar fashion to conventional microscopes with a large area source, i.e. they are duals. Scanning microscopes with a point detector (confocal microscopes) image coherently. The two-point resolution was calculated for confocal microscopes with one or two annular pupils. The OTF for a confocal microscope with one annular pupil or two annular pupils of differing radii was presented. Both these papers use the term 'confocal microscope'. These imaging schemes were the subject of a patent [7]. Cox *et al.* presented the imaging performance (two-point resolution and OTF) of confocal fluorescence microscopes, with circular or annular pupils [61].

Mahajan calculated Zernike polynomials for annular apertures [62, 63].

Diffraction by ellptical annuli has been considered in several papers [64-67].

A scalar zero-order Bessel beam generated by an angular spectrum of plane waves propagating along a cone of convergence angle $\theta$ can be written in cylindrical coordinates $\rho, \phi, z$

$$U = \exp(ikz\cos\theta)J_0(k\rho\sin\theta), \qquad (3)$$

where the phase at the origin is taken as zero. Note that this equation, unlike Linfoot and Wolf's results [6], is valid even for highly convergent beams, i.e. it is a solution of the Helmholtz equation. This differs from Gaussian beams, which are fundamentally a paraxial concept. It is also valid for arbitrary axial distance from the focal plane $z=0$. The zero-order Bessel beam is thus

$$U = \exp(ikz)\exp\left(-\frac{ikzs^2}{2}\right)J_0\left[k\rho s\left(1-\frac{s^2}{4}\right)^{1/2}\right], \qquad (4)$$

where $s = 2\sin(\theta/2)$. Then the paraxial limit, where $s$ is small, with optical coordinates $v, u$ given by $v = k\rho s, u = kzs^2$, giving approximately,

$$U = \exp(ikz)\exp\left(-\frac{ikzs^2}{2}\right)J_0(k\rho s) = \exp(ikz)\exp\left(-\frac{iu}{2}\right)J_0(v). \qquad (5)$$

For scalar waves, the amplitude in the focal region of an aplanatic, high numerical aperture, lens with a filter $P(\theta)$ in its front focal plane can be calculated by integrating over Bessel beams, to give

$$U = \exp(ikz)\int_0^\alpha P(\theta)\cos^{1/2}\theta J_0(k\rho\sin\theta)\exp(ikz\cos\theta)\sin\theta d\theta. \qquad (6)$$

Then for the paraxial case, again integrating over Bessel beams,

$$U = \exp(ikz)\int_0^1 P(p)J_0(vp)\exp\left(-\frac{iup^2}{2}\right)p\,dp, \qquad (7)$$

Where $p = s/[2\sin(\alpha/2)]$, $v = k\rho\sin\alpha$, and $u = 4kz\sin^2(\alpha/2)$.

For a plain annular pupil, $P(p) = 1$, $\varepsilon < p < 1$, this integral can be evaluated analytically in terms of Lommel functions [6], but this is beyond the scope of the present paper. However, near the optical axis, when $v \ll 1$, $v/u \ll 1$, the amplitude, normalized to unity at the focal point, becomes approximately

$$-\frac{i2\exp(ikz)}{(1-\varepsilon^2)u}\left[\exp\left(-\frac{iu\varepsilon^2}{2}\right)J_0(\varepsilon v) - \exp\left(-\frac{iu}{2}\right)J_0(v)\right]. \tag{8}$$

According to Young's edge diffraction theory, equivalent to the method of stationary phase, this can be recognized as a combination of approximations to two Bessel beams, given by scattering at the outer and inner rims of the aperture. For a plain circular aperture, $\varepsilon = 0$, so

$$U = -\frac{i2\exp(ikz)}{u}\left[1 - \exp\left(-\frac{iu}{2}\right)J_0(v)\right], \tag{9}$$

whereas for a narrow annulus, $\varepsilon \approx 1$,

$$U = \exp(ikz)\exp\left[-\frac{iu(1+\varepsilon^2)}{4}\right]\frac{\sin\left[\frac{u}{4}(1-\varepsilon^2)\right]}{\frac{u}{4}(1-\varepsilon^2)}J_0(v). \tag{10}$$

The focal spot is thus stretched out along the optical axis. In the limit as $\varepsilon \to 1$, the axial intensity is constant, corresponding to a true Bessel beam, but this is an idealization as a narrow annulus spreads light very quickly so the lens aperture would have to be much larger than the annulus. Truncation will result in fringes in the axial intensity along the optical axis.

An aplanatic nonparaxial system, or a system with an amplitude filter, changes the relative strength of the two contributions, so that destructive interference is no longer complete.

Both cases, of an annular pupil, with or without truncation by the lens aperture, differ from the Poisson spot for diffraction by a circular disk, when the axial intensity in the Fresnel region is constant, with a $J_0$ spot whose width is proportional to axial distance. On the other hand, diffraction by an annular mask gives an axial intensity that oscillates as a result of interference between the inner and outer boundaries of the annulus [10].

## 4 Axicons

McLeod showed that a conical prism can result in an axial line focus, with applications as an autocollimator or alignment device [10, 11]. According to McLeod, 'the name axicon means axis image', and a conical prism, in particular, is called a linear axicon. So the Arago spot is an example of an axicon, by this definition. McLeod discussed using diffraction by an annular aperture to form images along the optical axis (Fig. 2). This can be considered as an example of self-imaging, although the phenomenon of self-imaging had not been introduced at that time. He also showed that an axial line focus can be formed by reflection from the inside of a cylindrical tube (Fig. 3). We experimented with this approach in Oxford in the 1970's, but found it very sensitive to surface finish. It also has undesirable polarization properties, as discussed in Section 10 [4, 68].

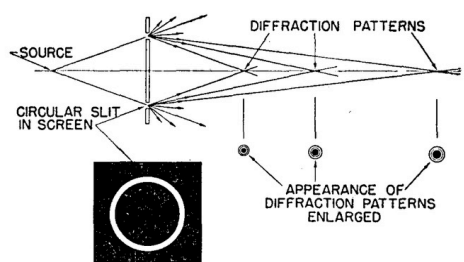

Fig. 1. Ray diagram showing formation of images of a point source by diffraction through a narrow circular opening.

Fig. 2. Producing an axial line focus using diffraction by an annular mask. (Reproduced from J. H. McLeod, *J. Opt. Soc. Am.* **44**, 592 (1954)), [10].

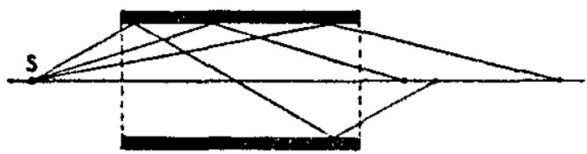

\
Fig. 3. Producing an axial line focus using reflection from the inside of a cylinder. (Reproduced from J. H. McLeod, *J. Opt. Soc. Am.* **44**, 592 (1954)), [10].

Rayces described that an axicon produces two images: a line image along the axis and a virtual annular image [69].

Kelly pointed out the equivalence of an annular pupil and an axicon [38].

Steel explained that a lens with negative spherical aberration can behave as an axicon in a zone of the lens, resulting in a 'lens axicon' [70]. In fact, Steel even went as far as to say that

'The essential property of an axicon is that it possesses spherical aberration.'

Use of aberrations to increase depth of focus has been described in many papers, but Steel's pioneering paper is rarely mentioned.

Dyson described the use of a circular grating, consisting of rings of constant radial spacing [12]. This is thus a diffractive axicon, analogous to the zone plate of ordinary focusing. He mentioned the earlier work of the Ronchi ruling [71]. He calculated that, ignoring edge effects, the grating produces a $J_0^2$ beam, and that 'the scale of the pattern is unchanged with axial position'. He also mentioned 'the pattern is achromatic as the wavelength does not appear in the argument of the Bessel function'. This latter point is discussed in Section pulsed. If the grating is tilted, astigmatism is introduced. He went on to consider spiral gratings, that produce higher-order Bessel vortex (doughnut) beams, $J_n^2$. Photographs were shown of both the zero and first order beams.

Fujiwara showed that a reflecting axicon also produces a Bessel beam [72].

Bickel; Bélanger and Rioux; and Perez *et al.* all described how an axicon in conjunction with a lens can be used to form an annular image [73-75]. An axicon produces a ring of light in the front focal plane of a lens. Such an annular image can then be used as the annular pupil of a further focusing lens, to efficiently produce a Bessel beam.

Bakken described a parabolic axicon, which he claims combines the effect of an axicon and a lens [76-78]. Bakken explains how an axicon, instead of focusing a point source on to an axial line, can perform the reverse operation of focusing a line source on to a point.

If a grating with transmission alternately zero and unity is used instead of an axicon, there is also a second diffraction order, where the rays are diffracted away from the axis. This corresponds to the twin image in holography, and can be eliminated by using a blazed grating or an annular stop [79, 80].

Perez *et al.* calculated the Fresnel diffraction pattern of a thin axicon, including along the axis, and in the ring focus [75].

## 5 Durnin, and publications in the 1990s

In 1987, Durnin *et al.* published two papers on Bessel beams [81, 82]. These papers are highly cited, with more than 8000 citations between them, and aroused considerable interest in the area of propagationally invariant beams. Perhaps this was a result of their use of the descriptions 'non-diffracting' or 'diffraction-free', which in our opinion is misleading, as Bessel beams are in fact a result of diffraction, the energy from the outer rings being diffracted inwards to the central lobe. Numerous papers were published in the 1990s, but many of these seemed unaware of the long history of the general principles of annular pupils, axicons or propagational invariance.

Indebetouw showed that the 3D pupil for a Bessel beam is a circle in $k$ space [83]. He pointed out the similarity with self-imaging. He proposed producing a Bessel beam using a Fabry-Perot cavity to select a longitudinal spatial frequency, and pointed out this process is achromatic.

Pieper studied two-pupil systems with annular apertures [84]. Two-pupil systems can be produced by interferometric methods, confocal imaging, or multi-photon effects.

Herman and Wiggins reviewed methods of generating Bessel beams [85]

Nakamura and Toyoda investigated optical imaging approaches to suppress the side lobes of the PSF with an annular pupil [86].

Cox and Anna investigated apodization of a Bessel beam, as in a Bessel-Gauss beam, but with an additional sharp-edged aperture [87]. Herman and Wiggins studied apodization of axicons [88].

Lü *et al.* and Bouchal *et al.* investigated truncation of Besel beams [89, 90]. Jaroszewicz discussed the connection between axicons and Bessel beams, and the effects of truncation [91].

Cox and Dibble produced a Bessel beam by spatially filtering light from a confocal cavity [92].

Jackson *et al.* calculated the 3D OTF for annular lenses with spherical aberration and defocus [93].

Strand studied the effect of combined aberrations on the Strehl ratio, both theoretically and experimentally [94].

Kowarz and Agarwal considered partially coherent Bessel beams [95].

Rosen presented a new family of pseudo-propagationally invariant beams, generated by an on-axis holographic filter [96, 97]. The simplest of these filters is basically an apodized fourth-power phase variation (radial harmonic), similar to spherical aberration [70].

Amidror investigated the Fourier transform of a circular cosine function, and described its behaviour in terms of half-order derivatives [98].

Herman and Wiggins calculated the beam propagation factor $M^2$ for a Bessel-Gauss beam [99].

Ruschin and Leizer investigated evanescent Bessel beams [100].

## 6 Annular pupils in confocal imaging systems

Annular pupils can be useful in confocal systems as the sidelobe levels of a Bessel beam are reduced [1-3, 7]. Sheppard and Gu calculated and plotted the three-dimensional (3D) OTF for an annular pupil [101]. An annular pupil results in an increased depth of focus, equivalent to a decreased axial resolution. But Gu and Sheppard found that for a confocal microscope with an annular illumination pupil and a finite-sized pinhole, according to geometrical optics, the signal from a mirror object goes to zero at a certain defocus distance, so that the optical sectioning effect is actually stronger than with a circular pupil [102, 103].

In fact, this type of behaviour had been noted earlier by Linfoot and Wolf, who observed [6]

'When the central obstruction is fairly large ... the pencil of rays passing through the geometrical focus has the form of a hollow circular cone ... and at a sufficient distance from the geometrical focal plane almost all the light is to be found between the walls of this cone.'

This prediction was found to hold according to diffraction theory, and was demonstrated experimentally [104-106]. Similar effects were also observed in fluorescence imaging [107-109]. It was shown that the missing cone of the 3D OTF that appears for a finite-sized pinhole was absent with an annular pupil. For a given value of pinhole radius, there is an optimum value of pupil obscuration $\varepsilon$.

Arimoto *et al.* reported a laser scanning microscope with axicon illumination [110]. They used both a glass axicon, and a blazed diffractive axicon [79].

Gu and Sheppard calculated the 3D OTF for a confocal microscope with two unequal annular pupils [106, 111], and Gan *et al.* calculated it for a confocal microscope with one annular and one circular pupil [112]. An important property to note is that for a narrow annulus and a circular pupil of approximately equal radius, the OTF depends critically on the relative size of the pupils: if the radius of the circle is equal to the inner radius of the annulus, the value of the OTF at zero spatial frequency is zero. Alternatively, if the circle is equal to the outer radius of the annulus, the OTF drops quickly to a value of 1/2 as spatial frequency is increased slightly from zero. These effects obviously greatly affect the contrast of a resulting image, and an optimum geometry is obtained for the intermediate case. They were anticipated for the case of two unequal narrow annuli many years earlier [3].

Gu *et al.* showed experimentally that an annular pupil can improve confocal imaging through a scattering medium [113].

Gan *et al.* calculated the Fresnel diffraction pattern of an annular aperture, and investigated the effect of altering the axial position of an annular pupil relative to the front focal plane of a lens, showing that the axial response becomes distorted [114].

**7 Higher order Bessel beams**

Higher order Bessel solutions were described in early papers by Lord Rayleigh, by Stratton and by Dyson [12, 115, 116].

Vasara et al. described generation of general Bessel beams with computer generated holograms [13]. Vortex beams can be generated by on-axis holograms with a spiral structure, and by off-axis holograms with a fork geometry [117].

Eqn. 4 was derived for a uniform summation over plane waves. In a similar way, higher order paraxial Bessel beams can be written as vortex beams,

$$U = \exp(ikz)\exp\left(-\frac{ikzs^2}{2}\right)J_n\left[k\rho s\left(1-\frac{s^2}{4}\right)^{1/2}\right]\exp(\pm in\phi), \quad (11)$$

or as multilobed beams,

$$U = \exp(ikz)\exp\left(-\frac{ikzs^2}{2}\right)J_n\left[k\rho s\left(1-\frac{s^2}{4}\right)^{1/2}\right]\begin{cases}\cos n\phi \\ \sin n\phi\end{cases}. \quad (12)$$

**8 Bessel-Gauss beams**

The Bessel-Gauss (BG) beam was first introduced by Sheppard and Wilson in 1978 [5], and then again, independently, by Gori *et al.* in 1987, soon after Durnin's papers [118]. Both papers give an analytic expression for the zero order BG beam, and plot its important properties. The beam is a solution of the scalar paraxial wave equation. In the focal plane, the amplitude is given by the product of a Bessel function and a Gaussian, thus avoiding the infinite energy problem of the Bessel beam. In the far field, the amplitude, apart from a quadratic phase term, is given by the convolution of a $\delta$-ring and a cylindrical Gaussian, which, if the Gaussian is much smaller than the ring is just a ring with a Gaussian cross section.

The BG beam in free space has two real parameters, which can be taken as the Gaussian confocal parameter, $z_0$, where $kz_0 \gg 1$, and the numerical aperture of the Bessel beam, $\sin\alpha \ll 1$. By appropriate choice of parameters, the solution degenerates into either a Gaussian beam ($\sin\alpha = 0$) or a Bessel beam ( $kz_0 \to \infty$). Following a paper by Youngworth and Brown, some published papers assume an arbitrary choice for one parameter, the ratio of the widths of the Bessel function and the Gaussian [119].

Ref. [5] gave an expansion for the BG beam in terms of standard Laguerre-Gaussian (LG) beams. Both Sheppard and Wilson, and Gori *et al.* expressed the BG beam in terms of a Bessel function of complex amplitude, but with real and imaginary Gaussian functions [5, 118]. An alternative approach is to express the Gaussian also as a function of complex argument [120, 121]. Then the amplitude of a Gaussian beam is, in cylindrical coordinates $\rho, z$,

$$U_G = \frac{\exp(ikz)}{1+i\zeta}\exp\left[-\frac{k\rho^2}{2z_0(1+i\zeta)}\right], \tag{13}$$

where $\zeta = z/z_0$. The Gaussian function then describes both the Gaussian fall-off in amplitude and the parabolic phase front, while the denominator $1+i\zeta$ gives the Gouy phase as well as the inverse square law in intensity. Introducing the parameter $a = z_0 \sin\alpha$, the zero-order BG beam is

$$U_{BG} = U_G \exp\left[-\frac{ika^2\zeta}{2z_0(1+i\zeta)}\right]J_0\left[\frac{ka\rho}{z_0(1+i\zeta)}\right] = \frac{\exp(ikz)}{1+i\zeta}\exp\left[-\frac{k(\rho^2+ia^2\zeta)}{2z_0(1+i\zeta)}\right]J_0\left[\frac{ka\rho}{z_0(1+i\zeta)}\right]. \tag{14}$$

The depth of focus is inversely proportional to $(kz_0 \sin^2\alpha)^{1/2}$.

In the far field, the Bessel function $J_0$ of imaginary argument is equal to $I_0$, the modified Bessel function:

$$U_{BG} = -\frac{iz_0}{z}\exp(ikz)\exp\left(\frac{ik\rho^2}{2z}\right)\exp\left[-\frac{kz_0(\rho-a\zeta)^2}{2z^2}\right]\left\{\exp\left(-\frac{ka\rho}{z}\right)I_0\left(\frac{ka\rho}{z}\right)\right\}. \tag{15}$$

The first decaying exponential is a shifted Gaussian, with maximum at $\rho = az/z_0 = \sin\alpha$, representing a Gaussian-weighted annulus. The quantity in braces is approximately constant over the domain where the decaying Gaussian is appreciable, i.e. $\rho \approx a\zeta$.

Then, more generally, applying the asymptotic expansion of the modified Bessel function with argument of large modulus (not valid near the axis), we obtain

$$U_{BG} = \frac{\exp(ikz)z_0^{1/2}}{[i2\pi ka^2\zeta(1+i\zeta)]^{1/2}}\exp\left[-\frac{k(\rho-a\zeta)^2}{2z_0(1+\zeta^2)}\right]\exp\left(-\frac{ika^2}{2z}\right)\exp\left[\frac{ik\zeta(\rho+a/\zeta)^2}{2z_0(1+\zeta^2)}\right]. \tag{16}$$

For large $z$, the last complex exponential represents the Gaussian parabolic phase front. For intermediate values of $z$, it represents a toroidal phase front, with its center of curvature at $\rho = -a/\zeta$. For $ka \gg (2kz_0)^{1/2}$, the radius of the annulus is large compared with the width of the Gaussian, and the toroidal phase is approximately linear over the cross section of the annulus, with a phase gradient $ka/z_0 = k\sin\alpha$, independent of the value of $z$, as long as it is large enough for the asymptotic expansion to be valid. The geometric focus of this annular phase gradient is found to coincide with the origin, $z = 0$, as shown in Fig. 4. This explains how a linear axicon placed in the Fresnel region produces an approximation to a BG beam [122]. The annular region exhibits a second phase derivative $k\zeta/[z_0(1+\zeta^2)]$ (the same as for a Gaussian beam), so the phase variation matches a parabolic axicon rather than a linear axicon. The phase gradient in the annular region can also be considered to match that of a lens with negative spherical aberration, illustrating the fact that spherical aberration can act like an axicon [70]. In the far field,

$$U_{BG} = -i\frac{\exp(ikz)z_0^{3/2}}{[2\pi k]^{1/2}az}\exp\left(\frac{ik\rho^2}{2z}\right)\exp\left[-\frac{k(z_0\rho-az)^2}{2z_0z^2}\right]. \tag{17}$$

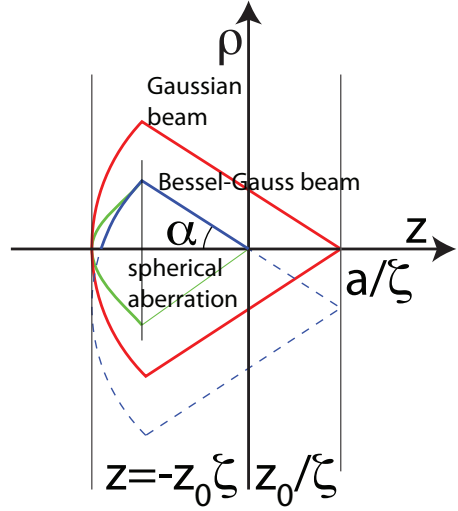

Fig. 4. A phase front of a Bessel-Gauss beam (in blue) compared with that of a Gaussian beam (in red). The ray at an angle $\alpha$ to the axis goes through the focal point. A spherically aberrated beam (in green) is also shown.

The Bessel-Gauss beam can be considered as an integral over Gaussian beams with propagation direction along the surface of a cone [5, 118], represented in the complex source point model as a summation of the fields of tilted complex sources [121].

Bagini *et al.* extended the Bessel-Gauss treatment to higher order beams [123]. They also considered the modified Bessel-Gauss beam, a dual beam, with near-field and far-field interchanged, also investigated later [124, 125].

Caron investigated beams where the Bessel function has a quadratic radial dependence [126].

The similarity of the annular-shaped far field of the eLG beam of high transverse order and the BG beam was observed by Saghafi and Sheppard [127]. They pointed out that the BG beams can be considered as approximations to the eLG beams, and *vice versa*. An important significance of eLG beams is their connection with multipole sources. However BG beams have the great advantage that they depend on a continuous, rather than a discrete, parameter, which may prove useful for fitting experimental data. The equivalence relationship for $n \gg l$ was given by Porras *et al.* [128]:

$$\rho^l L_n^{(l)}\left(\rho^2\right)\exp\left(-\rho^2\right) \approx n^{l/2} \exp\left(-\frac{\rho^2}{2}\right) J_l\left(2\sqrt{n}\rho\right). \tag{18}$$

Note the different Gaussian functions on the two sides. They proposed that an approximation to a propagationally invariant beam of BG form can be generated experimentally by using a Gaussian spatial filter on the waist of an eLG beam.

The general eLG beam can be written

$$U_{eLG} = \frac{\exp(ikz)}{(1+2i\zeta)^{n+l/2+1}} \left[\frac{k\rho^2}{z_0(1+2i\zeta)}\right] L_n^{(l)}\left[\frac{k\rho^2}{z_0(1+2i\zeta)}\right] \exp\left[-\frac{k\rho^2}{z_0(1+2i\zeta)}\right] e^{il\phi}, \tag{19}$$

where $z_0$ and $\zeta$ are defined for the BG beam, and we note that the associated Laguerre polynomial $L$ has the same complex argument as the Gaussian.

However, the amplitude can be written in the form

$$U_{eLG} = \frac{\exp(ikz)}{(1+2i\zeta)^{n+l/2+1}} \left[\frac{k\rho^2}{z_0(1+2i\zeta)}\right]^{l/2} L_n^{(l)}\left[\frac{k\rho^2}{z_0(1+2i\zeta)}\right] \exp\left[-\frac{k\rho^2}{z_0(1+4i\zeta^2)}\right] \exp\left[-\frac{2ik\rho^2\zeta}{z_0(1+4\zeta^2)}\right] e^{il\phi}, \tag{20}$$

where it is seen that the relative scaling of the Laguerre polynomial to the decaying Gaussian changes with propagation distance $\zeta$.

For large $n$, away from the focal plane, the beam has a doughnut form, and so we can use the asymptotic expansion of the Laguerre polynomial for large argument, valid if $k\rho^2/2z \gg n(n+l)$, to give

$$U_{eLG} \approx \frac{(-1)^n}{n!} \frac{\exp(ikz)}{(1+2i\zeta)^{n+l/2+1}} \left[\frac{k\rho^2}{z_0(1+2i\zeta)}\right]^{n+l/2} \exp\left[-\frac{k\rho^2}{z_0(1+2i\zeta)}\right] e^{il\phi}, \quad (21)$$

and in the far field

$$U_{eLG} \approx \frac{(-1)^{2n+l+1}}{n!} \exp(ikz) \left(\frac{z_0}{2z}\right) \left[\left(\frac{kz_0\rho^2}{4z^2}\right)^{n+l/2} \exp\left(-\frac{kz_0\rho^2}{4z^2}\right)\right] e^{il\phi}. \quad (22)$$

This result agrees with that derived from a limiting case of a generalized LG beam [125], or, apart from the parabolic phase term, by Fourier transformation of the amplitude in the waist [128]. We recognize that the expression in square brackets is similar to a Gamma distribution function of $kz_0\rho^2/4z^2$, which tends to a shifted normal distribution for large $n$. Matching up the mean and variance for Gamma and normal distributions, and taking $k\rho^2 z_0/4z^2 \approx n+l/2+1$, we obtain

$$U_{eLG} \approx \frac{(-i)^{2n+l+1}}{n!} \left(n+\frac{l}{2}+1\right)^{n+l/2} \exp(ikz)\left(\frac{z_0}{2z}\right)\exp\left(\frac{ik\rho^2}{2z}\right)\exp\left\{-\frac{kz_0}{2z^2}\left[\rho - 2z\left(\frac{n+l/2+1}{kz_0}\right)^{1/2}\right]^2\right\} e^{il\phi}. \quad (23)$$

Comparing Eq.23 with Eq.15, we see that the eLG and BG beams are of similar shape for large values of $ka^2/4z_0 = n+l/2+1$. Further, the approximation in Eq.21 is valid if $z/z_0 \gg n(n+l)/[2(n+l/2+1)]$.

Porras assumed that $n \gg l$. A better approximation for larger values of $l$ has been given by Mendoza *et al.* [122, 129].

Bessel beams are solutions of the Helmholtz equation, but Gaussian beams are a paraxial construction. Seshadri developed non-paraxial correction terms for BG beams [130].

Generalization of BG beams to the nonparaxial domain has been studied, based on the complex source point theory of Gaussian beams [121, 131].

**9 Pupil filters**
Thompson described using a simple filter of two rings, where each has a constant complex transmission, can increase the depth of focus of a lens [15].

Hegedus proposed the use of superresolving filters in confocal imaging [132, 133]. Optimized pupil filters were generated, and experimental measurements showed that:
'We see that the confocal system with a superresolving pupil has imaging characteristics similar to those of an unaberrated conventional imaging system but with twice the resolution.'

Sheppard and Hegedus presented expressions for the transverse and axial widths of the intensity point spread function in terms of the moments of the pupil [134].

Ojeda-Castañeda and Gomez-Sarabia presented the optimum filter for minimizing the second moment of the axial intensity for a given throughput [135].

Fukuda optimized edge-enhancing phase-shifting masks for use with annular illumination [136]

Sieracki and Hansen found that leaky annular masks, where the central region is partially transmitting, can result in transverse superresolution while maintaining good axial resolution [137]. Sheppard used the theory presented in Ref. [134] to investigate further the leaky annular masks [138]

Sheppard discussed synthesis of pupil filters for desired axial behaviour by Laplace or Fourier transformation of the desired axial amplitude [139]. Filters based on Butterworth or Hamming windows were presented.

Xu *et al.* reported a filter design with three elements of amplitude $1, 0, -1$ that gives a flat axial response [140].

Cizmar and Dholakia used a complex mask designed as the Fourier transform of the desired axial amplitude, after Gaussian filtering, for annular or axicon Bessel beam generation, with a liquid crystal spatial light modulator [141].

Wang and Gan designed by optimization three-ring and seven-ring filters to give an axially flat intensity [16, 142, 143]. Wang *et al.* extended this approach to a radially-polarized beam [144].

Sheppard *et al.* considered the properties of general two-zone filters [145]. They found the analytic condition connecting the modulus and boundary radius for a binary filter with phase difference of $\pi$ to give a flat axial intensity, and some special cases that give approximations to a Bessel beam.

Sheppard *et al.* considered the properties of general three-phase filters. They gave the solutions for axially flat intensity for phases of $0, \pi$ and for equal areas [146].

Sheppard obtained analytic solutions for axially-flat intensity with up to five rings by solving sets of nonlinear simultaneous equations [147]. Sheppard and Mehta showed that the axial range can be increased by introducing a central dark region [148], giving results similar to those in Ref. [140].

## 10 Polarization properties

Rayleigh, in a remarkable paper about electromagnetic wave propagation through waveguides published in 1897, showed from Maxwell's equations that the modes were of either transverse magnetic (TM) or transverse electric (TE) type [149]. He found that for the $TM_{lx}$ modes (where the subscript $x$ refers to the direction of the dominant electric field) of circumferential order $l \geq 1$, in cylindrical coordinates, are Bessel beams:

$$E_\rho = (1-t^2)J_l'(k\rho\sin\alpha)\cos l\phi,$$
$$E_\phi = -(1-t^2)\left[\frac{nJ_l(k\rho\sin\alpha)}{k\rho\sin\alpha}\right]\sin l\phi, \quad (24)$$
$$E_z = -2it J_l(k\rho\sin\alpha)\cos l\phi,$$

while for $TE_{lx}$ modes

$$E_\rho = (1+t^2)\left[\frac{nJ_l(k\rho\sin\alpha)}{k\rho\sin\alpha}\right]\cos l\phi,$$
$$E_\phi = -(1+t^2)J_l'(k\rho\sin\alpha)\sin l\phi, \quad (25)$$
$$E_z = 0.$$

Here $t = \tan(\alpha/2)$, where $\alpha$ is the angle a ray makes with the optical axis. We have introduced an arbitrary weighting for the modes, which will simplify later expressions. We have also suppressed the longitudinal variation $\exp[i(kz\cos\alpha - \omega t)]$, the transverse field variation being propagationally invariant. Rayleigh was interested in the waveguide modes, so applied boundary conditions at the outer surface to give radial orders, corresponding to specific values of $t = \tan(\alpha/2)$ but the original solutions of Maxwell's equations are valid even in free space. If the wave guide becomes very large, a continuous variation in $t$ becomes possible. There are also other sets of modes $TM_{ny}$, $TE_{ny}$ given by rotating the given ones by $90°$ about the axis. Here the subscripts $x, y$ refer to the direction of the dominant electric field.

Yoshida and Asakura calculated the electric energy density of a lens of high numerical aperture with an annular pupil illuminated by a plane polarized wave [150]. Sheppard considered the case of a narrow annulus illuminated by linearly polarized light, and discussed the mode structure [4]:

'An alternative way of considering the field in the focal region is in terms of a superposition of elementary cylindrical harmonics. This approach is discussed by Stratton [116], p.355 and it is seen that the field produced in the focal region of an annular system consists of only the first-order term of the expansion, which is made up of two modes, one with transverse **E** and one with transverse **H**.'

From Eqs.23 and 33, we have for $l = 1$, after using Bessel function identities, for $TM_{1x}$,

$$E_\rho = \tfrac{1}{2}(1-t^2)[J_0(k\rho\sin\alpha) - J_2(k\rho\sin\alpha)]\cos\phi,$$
$$E_\phi = -\tfrac{1}{2}(1-t^2)[J_0(k\rho\sin\alpha) + J_2(k\rho\sin\alpha)]\sin\phi, \qquad (26)$$
$$E_z = -2itJ_1(k\rho\sin\alpha)\cos\phi,$$

and for $TE_{1x}$

$$E_\rho = \tfrac{1}{2}(1+t^2)[J_0(k\rho\sin\alpha) + J_2(k\rho\sin\alpha)]\cos\phi,$$
$$E_\phi = -\tfrac{1}{2}(1+t^2)[J_0(k\rho\sin\alpha) - J_2(k\rho\sin\alpha)]\sin\phi, \qquad (27)$$
$$E_z = 0.$$

In Cartesian coordinates, these become, for $TM_{1x}$,

$$E_x = [J_0(k\rho\sin\alpha) - (1-t^2)J_2(k\rho\sin\alpha)\cos 2\phi],$$
$$E_y = -(1-t^2)J_2(k\rho\sin\alpha)]\sin 2\phi, \qquad (28)$$
$$E_z = -2itJ_1(k\rho\sin\alpha)\cos\phi,$$

and for $TE_{1x}$,

$$E_\rho = [J_0(k\rho\sin\alpha) + (1+t^2)J_2(k\rho\sin\alpha)\cos 2\phi],$$
$$E_\phi = (1+t^2)J_2(k\rho\sin\alpha)]\sin 2\phi, \qquad (29)$$
$$E_z = 0.$$

For the paraxial case, when we can assume $t$ is small, the $x$ components of electric field are equal, while the $y$ components are of opposite sign. So if we find the average of these two components, we obtain an approximately linearly polarized wave. For non-zero $t$, we then have the $HE_{1x} = LP_{0x}$ mode,

$$E_x = [J_0(k\rho\sin\alpha) + t^2 J_2(k\rho\sin\alpha)\cos 2\phi],$$
$$E_y = t^2 J_2(k\rho\sin\alpha)]\sin 2\phi, \qquad (30)$$
$$E_z = -2itJ_1(k\rho\sin\alpha)\cos\phi,$$

which is the result of focusing linearly polarized light from a narrow annulus. As $t$ increases, the cross-components of field, especially the longitudinal component, increase in strength, thus increasing the strength of the outer rings and the size and eccentricity of the central lobe [4]. By a value of $60°$, the central lobe splits into two [4]. We can also examine values of $\alpha$ greater than $90°$, which result from focusing with a grazing incidence paraboloidal mirror. Then, as $\alpha \to 180°$, after renormalizing,

$$E_x = J_2(k\rho\sin\alpha)\cos 2\phi,$$
$$E_y = J_2(k\rho\sin\alpha)\sin 2\phi, \qquad (31)$$
$$E_z = 0,$$

The intensity now exhibits a zero on the axis, i.e. it is a doughnut beam [4]: 'As $\alpha \to \pi$ the geometry again becomes 'paraxial' but the energy density tends to ... $J_2^2(v)$. Now the energy density is zero at the geometrical focus, the distribution is again circularly symmetric, but the maximum energy occurs on a circle with centre at $v = 0$.'

This corresponds to subtracting the *TE* mode from the *TM* mode, equivalent to the $EH_{1x}$ mode, sometimes written as $HE_{-1x}$. The electric field exhibits a dipole-like polarization. An analogous effect was described for a reflective axicon (called a waxicon) [68]. This is an important effect to consider for focusing by grazing incidence mirrors.

In 1962, Shimoda proposed generating a $TM_0$ beam in a cylindrical resonator. He described that there was a longitudinal electric field component of $J_0$ form, and suggested using this for laser acceleration of particles [151]. This idea was taken further in 1983 by Fontana and Pantel [152], and later, Romea and Kimura proposed using an axicon for laser particle acceleration [153]. $TM_0$ corresponds to illumination of a narrow annular pupil with radial polarization. In the paraxial case, the beam is a doughnut, with a dark centre. But for larger values of numerical aperture, there is a longitudinal electric field, and when $t > 1$, the longitudinal field dominates, and the beam has a bright centre. On the other hand, $TE_0$ corresponds to illumination with azimuthal polarization, and gives a doughnut beam for any value of $t$. Both these modes are cylindrically symmetric, so there is only one of each of these modes. Also in 1983, Brittingham proposed his focus wave mode, a pulsed *TE* beam [154. Both TM and TE pulsed beams were also analyzed by Sezginer {Sezginer, 1985 #2170]. Mishra described a $TM_0$ Bessel beam [155]. Hall and coworkers analyzed the azimuthal Bessel-Gauss beam [156-159], and Youngworth and Brown investigated highly focused azimuthal and radial Bessel-Gauss beams [160]. Dorn *et al.* showed that focusing radially polarized light with an annular pupil creates a focal spot smaller than focusing circularly polarized light [161]. Sheppard and Choudhury extended this concept to Bessel beams, including generation using solid immersion lenses (SILs) [162]. Yew and Sheppard explored tight focusing of radially-polarized Bessel-Gauss beams [163].

For the case when $l = 0$, we obtain for $TM_0$ [115],

$$E_\rho = (1-t^2)J_1(k\rho\sin\alpha),$$
$$E_\phi = 0, \qquad\qquad (32)$$
$$E_z = 2it J_0(k\rho\sin\alpha),$$

while for the $TE_0$ mode

$$E_\rho = 0,$$
$$E_\phi = \pm J_1(k\rho\sin\alpha), \qquad\qquad (33)$$
$$E_z = 0.$$

A range of more complicated polarization distributions can be generated by summation of the modes [116]. This summation is equivalent to integration of an angular spectrum of plane waves, incident along the surface of a cone [164]. According to a generalization of McCutchen's construction, the field in the focal region is given by the three dimensional (3D) Fourier transform of a vectorial distribution on the surface of a sphere in $k$ space [165], and for focusing with a narrow annular pupil this gives a directed circle on the sphere [83]. Any variation around the circle can then be expanded into two Fourier series, of radial (*TM*) and circumferential (*TE*) terms, respectively.

Eqs.24, 33 gave the field variation for $TM_{1x}, TE_{1x}$. By adding $HE_{1x} \pm iHE_{1y} = LP_{0x} \pm iLP_{0y}$, we generate circular polarization, giving cylindrically symmetric beams, $CP_{0\pm}$. This corresponds to illuminating an annular pupil with circularly polarized light. Similarly, approximately circularly polarized, vortex forms of the cylindrically symmetric modes, $CP_{1\pm}$, can be generated, by summing $TM_0 \pm iTE_0$, corresponding to an annular pupil illuminated by circularly polarized light with a charge 1 phase vortex.

In Eqs.24, 33, the azimuthal field variation is written in trigonometric form. An alternative is to express the azimuthal variation in complex exponential form. Then there are again two sets of modes, corresponding to clockwise and anticlockwise vortex beams [116, 166]. Then for illumination of an annular pupil with a charge 1 phase vortex, we have $TM_{1v} = TM_{1x} \pm iTM_{1y}, TE_{1v} = TE_{1x} \pm iTE_{1y}$.

In Cartesian coordinates, these give, for $TM_{1v}$,

$$E_x = \tfrac{1}{2}(1-t^2)[J_0(k\rho\sin\alpha) - J_2(k\rho\sin\alpha)]\exp 2i\phi,$$
$$E_y = \tfrac{i}{2}(1-t^2)[J_0(k\rho\sin\alpha) + J_2(k\rho\sin\alpha)]\exp 2i\phi, \qquad (34)$$
$$E_z = -2itJ_1(k\rho\sin\alpha)\cos\phi,$$

and for $TE_{1v}$,

$$E_x = \tfrac{1}{2}(1+t^2)[J_0(k\rho\sin\alpha) + J_2(k\rho\sin\alpha)]\exp 2i\phi,$$
$$E_y = \tfrac{i}{2}(1+t^2)[J_0(k\rho\sin\alpha) - J_2(k\rho\sin\alpha)]\exp 2i\phi, \qquad (35)$$
$$E_z = 0.$$

Note that the form (rather than the strength) of the expressions for $TE_{1v}$ is independent of $t$. On the axis, the electric field is circularly polarized. The average of $TM_{1v}$ and $TE_{1v}$ then gives the approximately circularly polarized $HE_{1v} = CP_{0+}$ mode,

$$E_x = [J_0(k\rho\sin\alpha) + t^2 J_2(k\rho\sin\alpha)\exp 2i\phi],$$
$$E_y = i[J_0(k\rho\sin\alpha) - t^2 J_2(k\rho\sin\alpha)\exp 2i\phi], \qquad (36)$$
$$E_z = -2itJ_1(k\rho\sin\alpha)\cos\phi.$$

Bessel beams of different polarizations, including circularly polarized modes and higher order vortex beams, were analyzed by Bouchal and Olivíc [167].

Sick *et al.* proposed altering the polarization of annular illumination to generate a focused spot with arbitrary orientation in three dimensions [168].

The $HE_{1x} = LP_{0x}$ mode can alternatively written in a different basis, as the sum of a transverse electric dipole mode $p_x$ and a crossed magnetic dipole mode (oriented in the $y$ direction), $m_y$ [169, 170]. The polarization on the $k$ space circle is then the same as the polarization in the far field of the corresponding dipoles situated at the centre of the $k$ space sphere. Thus any combination of $TM_{1x}$ and $TE_{1x}$ modes can also be considered as a sum of $p_x$ and $m_y$.

In Cartesian coordinates, we then have, for $p_x$,

$$E_x = (1+t^4)J_0(k\rho\sin\alpha) + 2t^2 J_2(k\rho\sin\alpha)\cos\phi,$$
$$E_y = 2t^2 J_2(k\rho\sin\alpha)\sin\phi, \qquad (37)$$
$$E_z = -it(1-t^2)J_1(k\rho\sin\alpha)\cos\phi,$$

and for $m_y$,

$$E_x = (1-t^4)J_0(k\rho\sin\alpha)\cos\phi,$$
$$E_y = -(1-t^4)J_0(k\rho\sin\alpha)\sin\phi, \qquad (38)$$
$$E_z = -it(1+t^2)J_1(k\rho\sin\alpha)\cos\phi.$$

In cylindrical coordinates, for $p_x$,

$$E_\rho = \left[(1+t^4)J_0(k\rho\sin\alpha) + 2t^2 J_2(k\rho\sin\alpha)\right]\cos\phi,$$
$$E_\phi = -\left[(1+t^4)J_0(k\rho\sin\alpha) - 2t^2 J_2(k\rho\sin\alpha)\right]\sin\phi, \qquad (39)$$
$$E_z = -it(1-t^2)J_1(k\rho\sin\alpha)\cos\phi,$$

and for $m_y$,

$$E_\rho = (1-t^4)J_0(k\rho\sin\alpha)\cos\phi,$$
$$E_\phi = -(1-t^4)J_0(k\rho\sin\alpha)\sin\phi, \qquad (40)$$
$$E_z = -it(1+t^2)J_1(k\rho\sin\alpha)\cos\phi.$$

It is found that the electric dipole mode $p_x$ gives the lowest level of side lobes for any combination of $TM_{1x}$ and $TE_{1x}$ [169, 170]. It gives a smaller central lobe and lower side lobes than $LP_{0x}$, and than radial polarization, $TM_0$, for $\alpha \leq 60°$ [170]. However, the Bessel beam with the smallest central lobe is given by the $TE_{1x}$ mode [170]. The cylindrically symmetric version of the $TE_{1x}$ mode is the azimuthally polarized vortex mode, $TE_{1v}$, which gives an even smaller central lobe [170, 171]. The cylindrically symmetric version of $p_x$, $p_x + ip_y$, also gives low side lobe level [170, 172]. This corresponds to a rotating electric dipole field $p_{rot}$, equivalent to an annular pupil illuminated with elliptically polarized light with major axis in the azimuthal direction, with a charge 1 phase vortex.

For circularly polarized illumination of an annular pupil, the intensity, defined as the time-averaged electric energy density and normalized to unity at $v = 0$, where $v = k\rho\sin\alpha$, is [162]

$$I(v) = J_0^2(v) + 2t^2 J_1^2(v) + t^4 J_2^2(v). \qquad (41)$$

For radially polarized illumination of an annular pupil ($TM_0$), the intensity is [162]

$$I(v) = J_0^2(v) + \frac{(1-t^2)^2}{4t^2} J_1^2(v). \qquad (42)$$

For azimuthally polarized illumination with a phase vortex $TE_{1v}$, the intensity is [170, 171]

$$I(v) = J_0^2(v) + J_2^2(v). \qquad (43)$$

For a vortex electric dipole polarization ($p_{rot}$), the intensity is

$$I(v) = J_0^2(v) + 2\frac{t^2(1-t^2)^2}{(1+t^4)^2} J_1^2(v) + \frac{4t^4}{(1+t^4)^2} J_2^2(v). \qquad (44)$$

The intensity profiles are compared in Fig. 5 for the case when the numerical aperture is 0.95. The width of the intensity distribution is 10.59% larger than the Airy disk for the circularly polarized case $CP_0$, at a numerical aperture of 0.95. It is 27.95% narrower than the Airy disk for the radially-polarized case $TM_0$, 28.25% narrower for azimuthally polarized with a vortex $TE_{1v}$, 25.15% narrower for $p_{rot}$, and 30.31% narrower for a scalar $J_0$ Bessel beam.

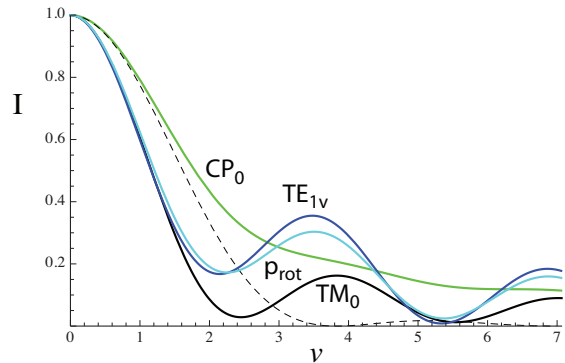

Fig. 5. The normalized intensity profile across Bessel beams produced by focusing light by a lens of numerical aperture 0.95: radially polarized $TM_0$, circularly polarized $CP_0$, azimuthally polarized with a vortex $TE_{1v}$, rotating electric dipole $p_{rot}$. The profile for an Airy disk is shown (dashed line) for comparison.

The cross-sections of the non-cylindrically symmetric beams can be compared by considering the area of the central lobe in the parabolic approximation, when the parabolic cross-section is elliptical. These are shown plotted against numerical aperture in Fig. 6. To give some appreciation of the magnitudes of these areas, a spot with FWHM of $\lambda/2$ gives an area/$(\lambda/n)^2$ of 0.196, the parabolic approximation for a scalar Airy disk with $NA/n = 1$ gives an area/$(\lambda/n)^2 = 0.159$, and the parabolic approximation for a scalar Bessel beam with $NA/n = 1$ gives an area/$(\lambda/n)^2 = 0.080$. The case for $TM_{1x}, TM_{1v}$ are not shown as there is a saddle or minimum intensity at the centre. It is seen that $TE_{1v}$ gives the smallest area, becoming equal to that of $TM_0$ for $NA = n$.

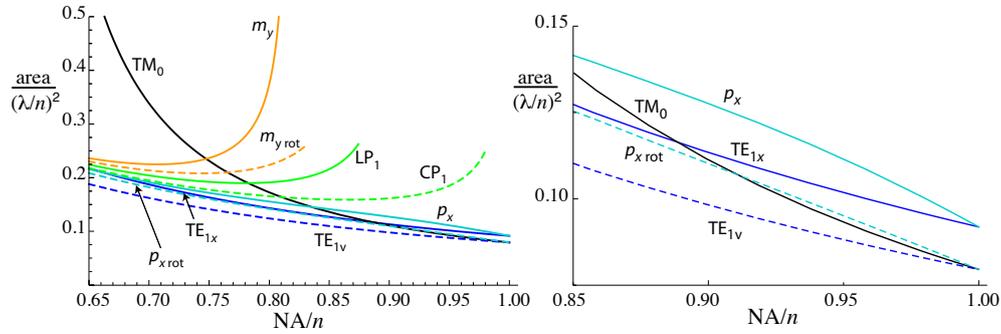

Fig. 6. The area of the parabolic central lobe of a Bessel beam generated by focusing different polarized illuminations. Dashed lines refer to rotating fields (circular polarized, rotating dipoles or azimuthal vortex. The right-hand plot shows an enlarged view for high NAs.

Focusing of Bessel beams for arbitrary orders of polarization and phase vortices has also been analyzed [173]. An interesting observation is that there are solutions for a half order polarization vortex combined with a half order phase vortex [173].

## 11 Waveguide and resonator modes

The solution of the wave equation in cylindrical coordinates leads to expressions for the electric and magnetic fields in terms of Bessel functions of the radial coordinate and trigonometric (or complex exponential) functions in the circumferential coordinate. As we have mentioned, in 1897 Lord Rayleigh gave the solution for electromagnetic waves propagating inside a perfectly conducting cylindrical tube, with either rectangular or circular cross-section [115]. He showed that these modes were either of TE or TM type, and pointed out that there was a cut-off frequency for all the modes. For a circular waveguide, the modes are written $TE_{ln}$, $TM_{ln}$, where for TE $l$ is the order of the Bessel function for $H_z$, $E_\rho$ and $H_\phi$, and for TM for $E_z$, $E_\phi$ and $H_\rho$. Plots of the field distributions for the 30 lowest order modes have been presented by Lee *et al.* [174]. The dominant mode (that with the lowest cut off frequency) is $TE_{11}$, followed by $TM_{01}$, $TE_{21}$, and then $TE_{01}$ and $TM_{11}$ with the same cut off.

Hondros and Debye considered the modes of a dielectric waveguide, and found that, apart from the $TE_{0n}$ and $TM_{0n}$ modes, the modes have both electric and magnetic $z$ components [175]. These hybrid modes are called $EH_{ln}$ and $HE_{ln}$ modes [176, 177]. Here the subscript $n$ refers to the order of the Bessel function, i.e. the azimuthal mode number, and $m$ refers to the radial mode number. The $HE_{11}$ mode has no cut off, and the transverse electric field is in the $x$ or $y$ direction. The $TE_{01}$ and $TM_{01}$ modes, with the same cut off, have circumferential or radial transverse electric field, respectively. The $HE_{12}$ and $EH_{11}$ modes have the same cut off: $HE_{12}$ has transverse electric field in the $x$ or $y$ direction, but for $EH_{11}$ are of dipole shape. The $HE_{21}$ and $EH_{21}$ modes have transverse electric field loci of hyperbolic shape. Gloge showed that in the limit for small refractive index change (weakly guided modes), there is a redundancy so that linearly polarized modes $LP_{ln}$ are solutions [178]. The $LP_{01}$ mode is equal to the $HE_{11}$ mode, which has no cut off. $LP_{1n}$ is a combination of $HE_{2n}$, $TM_{0n}$ and $TE_{0n}$ modes. For $(l>1)$, $HE_{(l-1)n}$ and $EH_{(l+1)n}$ combine to give $LP_{ln}$. These results were anticipated by Snitzer [176], who said 'For the hybrid modes $(l \geq 1)$ the field distribution simplifies considerably for the case of a small index difference between the core and the cladding.'

McCumber investigated the effect of annular mirrors, i.e. of a circular output coupling aperture, on the modes of a laser resonator [179]. The power losses for the lowest order $TEM_{00}$ mode become larger in comparison to the next lowest, $TEM_{10}$, mode. Nakasuji and Ohtsuka have shown that if an annular aperture around the centre of output mirror is used for coupling out the laser energy, the lowest power loss among the resonant modes is for the lowest order transverse mode [180]. Annular resonators are often used as unstable resonators [46, 181, 182].

An annular resonator using toroidal mirrors, for which Bessel-Gauss beams are the modes, was proposed by Sheppard and Wilson [5]. A symmetrical confocal annular resonator is illustrated in Fig. 7. Related structures were described by Uehara and Kikuchi, Jabczynski, Pääkkönen and Turunen, and Rogel-Slazar *et al.* [183-185].

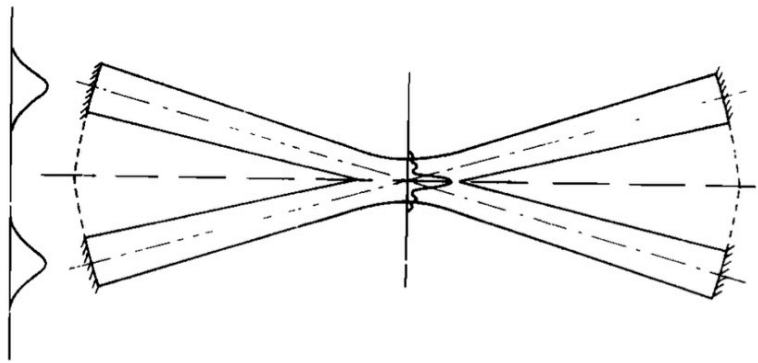

Fig. 7. A confocal annular resonator with toroidal mirrors. After Sheppard and Wilson [5].

## 12 Polychromatic beams and pulsed beams

For the same angular spectrum, blue light is focused by a lens to a smaller spot than red light (Fig.7, left). For beams with waists of the same cross section, red light diffracts more than blue light (Fig. 7, right). If, on the other hand, we have a spherical1y symmetric white light source, then by considerations of symmetry and conservation of energy, the spectral distribution must remain invariant. It is interesting, therefore, that a planar point source thus behaves differently than a spherical point source [186].

As pulsed beams can be considered as a coherent composition of spectral components, it is thus necessary to define the relationship between monochromatic components of different wavelength.

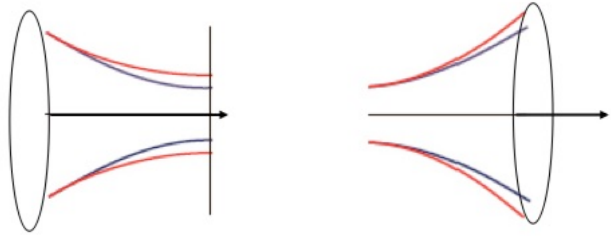

Fig. 8. Left: For the same angular spectrum, blue light is focused to a smaller spot than red light. Right: For a waist of the same cross section, red light diffracts more than blue light.

So far we have considered monochromatic Bessel beams. Dyson considered illumination of a diffractive axicon with polychromatic light [12]. He pointed out that the width of the Bessel beam produced has a width independent of wavelength. Rather than a polychromatic beam, where the different spectral components are summed incoherently, we can also consider the case of a pulsed beam, where the spectral components are added coherently, as studied by Campbell and Soloway [187]. There is, of course, an intermediate case where the spectral components are added partially-coherently. A pulsed beam generated using a diffractive axicon, we call a Type 1 pulsed beam. It has the property that the transverse wave number $k_\rho$ is independent of wavelength, corresponding to the third part of Fig. 1. A similar Type 1 pulsed Gaussian beam was analyzed by Christov [188], and by Liu and Fan [189]. It has spectral components that have the same beam waist, but varying confocal parameter [186].

Burckhardt described a pulsed acoustic imaging system using a refractive axicon [190]. In this case, the angular spectrum in the far field, neglecting material dispersion, is independent of wavelength. Such a beam was later called by Lu and Greenleaf an X-wave [191, 192]. We call this a Type 2 pulsed beam. It corresponds to the second part of Fig. BesselBeams. Fagerholm *et al.* calculated the field of an X-wave, by integrating over Bessel beams with constant angle of convergence with a Gamma spectral distribution [193]. They also considered higher order vortex X-waves. X-waves were observed experimentally by Saari and Reivelt, and Sonajalg *et al.* [194, 195]. Electromagnetic (TE or TM) X-waves were investigated theoretically by Recami [196]. The corresponding Type 2 pulsed Gaussian beam was investigated by Federico and Martinez [197]. It has spectral components that have constant ratio of confocal parameter to beam waist [186].

An intermediate case (Type 3) has spectral components that are scaled the same in the near and far field. In a paraxial theory, they have the same phase, so the spectral components propagate in step. Such pulses propagate at the speed of light, and were introduced by Brittingham [154]. Brittingham called the solutions focus wave modes (FWMs). Bélanger showed that the Helmholtz equation reduces to a paraxial wave equation, and derived elegant Laguerre-Gaussian (eLG) solutions [198]. Sezginer explained that focus wave modes are nonphysical (infinite energy, like the Bessel beam or a plane wave) and obtained an Hermite-Gaussian (HG) solution [199]. Overfelt obtained a pulsed Bessel-Gauss beam solution by assuming a spectral distribution given by the product of a modified Bessel function $I_0$ and a Gaussian [120]. Ziolkowski *et al.* presented an angular spectrum representation of FWMs, showing that they include backward propagating components [200]. Corresponding Type 3 pulsed Gaussian beams were described by Heyman (called isodiffracting pulses) and by Sheppard and Gan (called Gaussian pulse beams) [201, 202] [186]. Type 3 Gaussian beams propagate with a constant spectral content for points on the axis. They have spectral components that have the same confocal parameter, but varying beam waist [186]. This type of pulse also has the important property that the phase fronts for the different spectral components match up, so that it is the form of pulsed beam produced in a mode-locked laser cavity, which requires a particular relationship of the beam waists for different spectral components, regardless of the temporal pulse shape. In fact such a behaviour is well known for propagation of pulses in square law media, such as graded index optical fibres or in lens waveguides [186]. For the case of integration over Gaussian beams, we obtain a pulsed Gaussian beam, where the pulses propagate through a stationary Gaussian envelope.

Assuming a Gamma spectral distribution has the important property that it avoids artefacts from nonpositive frequency components, which is not the case for an assumed Gaussian spectral distribution. Sheppard and Gan, alternatively, assumed a Gaussian pulse in the time coordinate [186], giving a solution for pulsed Gaussian beams in terms of error functions. Sheppard derived a Type 3 pulsed Bessel beam, assuming a spectral distribution given by a Pearson Type III distribution (shifted Gamma distribution) [203], which can avoid low (as well as negative) frequency components. Again the solution can be expressed in terms of eLG functions. Reivelt and Saari assumed a spectral distribution given by a modified Bessel function and a Gaussian, giving a Bessel-Gauss solution [204].

For a summation over Bessel beams with a spectral distribution $f(k)$,

$$U = \int_0^\infty \exp[ikz\cos\theta(k)] J_0[k\rho\sin\theta(k)]\exp(-ikct) f(k) dk, \qquad (45)$$

where $\theta$ is a function of $k$. This can be written (as in Eq. 4)

$$U = \int_0^\infty \exp\left(-\frac{ikzs^2}{2}\right) J_0\left[k\rho s\left(1-\frac{s^2}{4}\right)^{1/2}\right] \exp(-ikct') f(k) dk, \quad 0 \le s \le 1, \qquad (46)$$

where $s = 2\sin(\theta/2)$ is in general a function of $k$, and the retarded time is $t' = t - z/c$. For Type 1 pulses, the argument of Bessel function is independent of $k$, and the Bessel function can be taken outside of the integral. For Type 2 pulses, $s$ is independent of $k$.

The Type 3 pulse condition is $ks^2 = 2k_c$, where $k_c$ is a positive constant. The minimum value of $k$ is set by the condition $s = 2$, giving $k_{min} = k_c/2$. Then the first exponential can be taken outside of the integral:

$$U = \exp(-ik_c z) \int_0^\infty J_0\left[k_c^{1/2}(2k - k_c)^{1/2}\rho\right] \exp(-ikct') f(k) dk, \qquad (47)$$

where the lower limit of the integral, corresponding to $s = 2$ ($\theta = \pi$), is now taken as $k_c/2$. Along the axis, the integral is simply the Fourier transform of the spectral distribution.

However, if we are trying to obtain an analytic solution, it may be preferable to ensure the arguments of the Bessel function and the exponential inside the integral are similar, and to change the variable of integration:

$$\begin{aligned} U &= \exp\left[\frac{-ik_c(z+ct)}{2}\right] \int_{k_c/2}^\infty J_0\left[k_c^{1/2}(2k-k_c)^{1/2}\rho\right] \exp\left[-\frac{i(2k-k_c)ct'}{2}\right] f(k) dk \\ &= \exp\left[\frac{-ik_c(z+ct)}{2}\right] \int_0^\infty J_0\left[(2k_c k')^{1/2}\rho\right] \exp(-ik'ct') f\left(k' + \frac{k_c}{2}\right) dk', \end{aligned} \qquad (48)$$

where $k' = k - k_c/2$. We observe the backward propagating phase term before the integral, representing a luminal, forward propagating, propagationally-invariant, localized pulse envelope. Then assuming an exponential spectral distribution leads to the original FWM solution, and assuming a Pearson Type III distrubution leads to the eLG solutions. Assuming a modified Bessel distribution, similar in form to a Rice distribution, leads to a Bessel-Gauss distribution [120, 204]. As we mentioned earlier, the similarity of high order eLG beams to BG beams has been noted [127, 128].

Alternatively, we can impose a minimum wave number $k_{min}$, and introduce $k'' = k - k_{min}$, to give

$$U = \exp\left[-ik_c z - ik_{min} ct'\right] \int_0^\infty J_0\left[k_c^{1/2}(2k'' + 2k_{min} - k_c)^{1/2}\rho\right] \exp(-ik''ct') f(k'' + k_{min}) dk''. \qquad (49)$$

In particular, if we put $k_{min} = k_c$, corresponding to just eliminating completely the backward propagating components, then the exponential in front of the integral becomes $\exp(-ik_c t)$, i.e. independent of position. Sheppard and Saari showed that for a shifted exponential distribution, analytic solutions can be obtained in terms of Lommel functions for different integration limits [205]. For a shifted exponential spectral distribution

$$f(k) = z_0 \exp\left[-\left(k - \frac{k_c}{2}\right)\right], \quad k_1 > k > \frac{k_c}{2}, \qquad (50)$$

from Eq.22 in Ref. [205], the amplitude can be written in the simple form

$$U_{k_c/2, k_1} = \exp\left[\frac{i}{2}(k_c z - 2k_1 ct)\right](2k_1 - k_c)\frac{1}{u}\left[U_1(u,v) + iU_2(u,v)\right]. \qquad (51)$$

Here $U_1, U_2$ are Lommel functions, and

$$\begin{aligned} u &= (2k_1 - k_c)(ct' - iz_0), \\ v &= [k_c(2k_1 - k_c)]^{1/2} \rho. \end{aligned} \qquad (52)$$

For the special case where $k_1 = k_c$, $u = k_c(ct' - iz_0)$, $v = k_c \rho$,

$$U_{k_c/2, k_c} = \exp\left[\frac{ik_c}{2}(z - 2ct)\right]k_c z_0 \frac{1}{u}\left[U_1(u,v) + iU_2(u,v)\right]. \qquad (53)$$

This corresponds to the case where there are only backward propagating components, called a truncated spectrum FWM [205]. In this solution, the pulse envelope travels at the speed of light in the direction opposite to all its spectral components, a free-space version of backward light. By subtracting two solutions with different values of $k_1$, analytic bandlimited FWM solutions can be obtained [205].

An angular spectrum can be represented as a surface in $k$ space. For focusing of continuous waves the surface is just the cap of a spherical shell, so that $k$ is independent of direction [165]. Focusing with a narrow annular pupil gives a circle on the sphere [83]. The amplitude in the focal region is then given by the 3D Fourier transform of this 3D pupil [165]. The 3D OTF is given by the autocorrelation of the 3D pupil. The angular spectrum for different types of pulsed Bessel beams is then as shown in Fig. 9. Type 1 beams have constant transverse $k$, so are represented as a cylinder in $k$ space. Type 2 beams (X-waves) have constant angle of convergence, so are represented as a cone in $k$ space. Type 3 beams (focus wave modes) have the equation $k_\rho^2 = k_c(k_c + 2k_z)$, and are represented as a paraboloid in $k$ space. Values of $k$ with $k_c/2 \le k \le k_c$ give backward propagating spectral components. The fact that the 3D pupil function is represented by a paraboloid is equivalent to the observation that focus wave mode solutions of the Helmholtz equation are products of a function that is a solution of the paraxial wave equation and a complex exponential phase term [198].

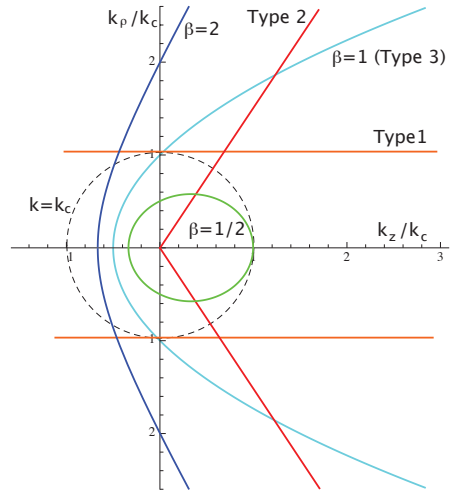

Fig. 9. A section through $k$ space for different types of pulsed beam. The dashed curve shows the case for focusing of monochromatic waves with $k = k_c$. $k_z > 0$ corresponds to forward propagating components.

Fig. 10 shows the dispersion relationships, $k$ versus $k_z$, for the different types of pulsed beam [122]. Plane waves propagating along the $z$ axis in free space are shown as dashed lines. Monochromatic focusing is given by constant $k$, with $-k \leq k_z \leq k$. Type 1 pulses are given by the hyperbola where $(k^2 - k_z^2)$ is a constant. Type 2 pulses are given by the line $k = k_z \cot\alpha$, where $\alpha$ is constant. Type 3 pulses are given by the line $k = k_z + k_c$. The slope is unity, representing a group velocity of $c$, whereas for Type 1, the group velocity is less than $c$, while for Type 2 it is greater than $c$. Type 1 pulses satisfy the relationship that the product of group and phase velocities is equal to $c^2$.

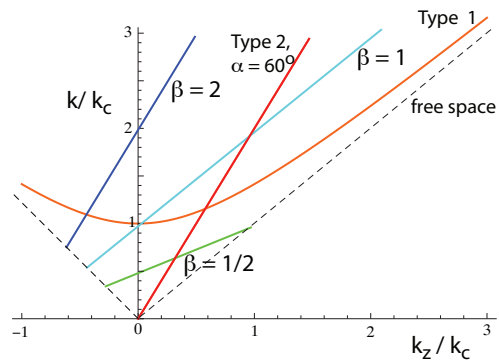

Fig. 10. The dispersion curves for different types of pulsed beam. The dashed lines show the case of a plane wave traveling along the $z$ axis. $k_z > 0$ corresponds to forward propagating components.

The transverse cross sections of the pulse for Type 3 (focus wave mode) and Type 2 (X-wave) are shown in Fig. 11 (a) and (b), respectively [122]. In both cases, although a Bessel beam has strong outer rings, these are much reduced for the pulsed beams, especially for the Type 3 case. The sidelobes are averaged out, but also tend to cancel out because of phase effects.

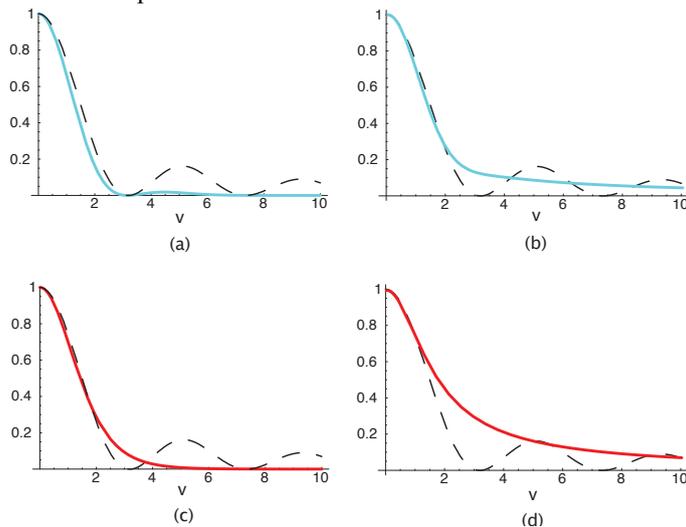

Fig. 11. Transverse cross sections through the intensity profile for different types of Bessel beams. (a) Pulsed beam, Type 3 case (focus wave mode). (b) Polychromatic continuous wave, Type 3 case (focus wave mode). (c) Pulsed beam, Type 2 case (X-wave). (d) Polychromatic continuous wave, Type 2 case (X-wave).

The focus wave mode (Type 3) solution gives a pulse that propagates at the speed of light. In a similar way, pulses propagating at $\beta c$, slower ($\beta < 1$), or faster ($\beta > 1$) than the speed of light can also be generated [206]. We choose the condition $k(1/\beta - \cos\theta) = k_c$. Then

$$U = \exp(-ik_c z)\int_{k_{\min}}^{k_{\max}} f(k) J_0\left\{\left[k^2\left(1-\frac{1}{\beta}\right)+\frac{2k_c k}{\beta}-k_c^2\right]^{1/2}\rho\right\}\exp(-ikct')dk, \qquad (54)$$

where now the local time is $t' = t - z/\beta c$. The lower limit is $k_{\min} = \beta k_c/(1+\beta)$. The upper limit for $\beta < 1$ is $\beta k_c/(1-\beta)$, but for $\beta > 1$ it is $\infty$. For $\beta > 1$, a Pearson Type III spectral distribution is therefore appropriate. The lower limit can be chosen to be $\beta k_c$ in order to eliminate backward propagating components completely. For the case when $\beta < 1$, the spectral distribution can alternatively be taken as a shifted beta distribution, which has bounded support and is also readily Fourier transformed [206].

Several papers have regarded backward propagating waves as noncausal. However, backward propagating waves are valid solutions of the Helmholtz equation, although they would limit long-distance communications applications. Backward propagating components are also a feature of the complex source point model of nonparaxial Gaussian beams. The connection stems from the observation that the FWM solution corresponds to the field of a traveling complex source, higher order solutions in terms of eHG or eLG modes being related to complex multipole sources [198, 203, 206].

The fact that the group velocity of localized wavepackets can be superluminal has raised the question of whether this phenomenon can allow information to be transmitted superluminally. Sauter and Paschke explained that it cannot [207]. They argued that the pulses do not actually travel along the axis. The pulses are an interference effect caused by luminal waves traveling at an angle to the axis, so information has to be sent off-axis even though it is detected on-axis. The misunderstanding of the superluminality follows directly from the erroneous view that the power in a Bessel beam travels along the axis.

Fischer *et al.* investigated the propagation of polychromatic, rather than pulsed, X-waves [208]. In Fig. 10 we compare the intensity cross sections for pulsed and polychromatic waves, of both Type 3 and Type 2. In both cases the outer rings of the Bessel beam are averaged out. The central lobe for the Type 2 case is broader, and the outer tail stronger. For the corresponding Type 1 case, the intensity cross section is independent of wavelength, and so the outer rings are maintained in the polychromatic case.

## 13 Nonlinear optics of Bessel beams

Wulle used Bessel beams to generate second harmonics in a crystal [209].

Glushko et al. studied harmonic generation by a Bessel beam [210].

Peatross et al. generated higher-order harmonics using a lens with annular aperture [211].

Peet and Tsubon generated third harmonics with a Bessel beam [212].

Tewari *et al.* presented a theory of third harmonic generation by Bessel beams [213].

Hell *et al.* demonstrated two-photon fluorescence microscopy using a lens with annular aperture [214].

Caron and Potvliege compared theoretically the generation of third harmonics by Bessel-Gauss beams and Gaussian beams [215].

Johannisson et al. investigated the effects of Kerr nonlinearity on Bessel beam propagation [216].

Botcherby et al. described two photon fluorescence microscopy using a Bessel beam, produced by a diffractive axicon lens combination [80]. Two high depth of focus images were used to generate stereo pairs.

## 14 1D beams, Airy beams, extended depth of focus, and light sheets

It is interesting to consider the case of one-dimensional (1D) propagation in space. Then the 1D analogy to the Bessel beam is simply a set of cosine fringes. The cosine fringes can be generated by diffraction from two narrow slits, analogous to an annulus in 2D. The axicon is the circularly symmetric version of the Fresnel biprism.

Berry and Balazs introduced the Airy wave-packet, a solution of the paraxial wave equation [217]. The Airy function is the Fourier transform of the complex exponential of a cubic phase variation. It exhibits strong fringing on one side of the peak value. The corresponding Airy beam in the spatial domain is a propagationally invariant structure, that appears to move in a parabolic path. While its peak moves in a parabolic path, the Airy beam carries infinite power, so its centre of gravity is undefined.

Rosen *et al.* introduced a 1D pseudo-propagationally invariant beam [218], analogous to their 2D pseudo propagationally invariant beam [96, 97]. Salik et al. demonstrated how 1D holographic elements for 1D beam shaping can be calculated by optimization [219].

Ojeda-Castañeda *et al.* used an apodizer to increase depth of focus, combined with digital restoration [220].

Sivilogiou *et al.* observed Airy beams experimentally [221].

More than ten years before that, Dowsky and Cathey used a lens with strong cubic phase in Cartesian coordinates to give a greatly increased depth of focus [222]. As the Airy function is the Fourier transform of the complex exponential of a cubic phase, the point spread function is essentially the same as for an Airy beam, so the image is poor until processed computationally. Arnison *et al.* used a similar approach for high numerical aperture imaging in a fluorescence microscope [223].

Vo *et al.* compared the 2D Airy beam with a beam aberrated with coma, and showed that an Airy beam can be considered as a beam with both coma and trefoil aberrations [224].

Sheppard obtained analytic solutions for maximally axially-flat intensity for the 1D case with binary filters with up to seven elements, by solving sets of nonlinear simultaneous equations [225]. These solutions give a smooth, flat axial intensity over a restricted range. Wilding *et al.* demonstrated this approach experimentally [226].

Sheppard showed that cosine-Gauss beams can be extended to the nonparaxial regime using the complex source point theory [121]. The complex source point approach avoids the introduction of evanescent waves.

## 15 Matthieu beams, Pearcey beams

Gutiérrez-Vega et al. introduced Mathieu beams, a solution of the Helmholtz equation in which for the zero-order solution the angular spectrum around a cone of illumination is modulated by an angular Mathieu functions $ce_0$ [227]. The amplitude in the propagation direction is similar in shape to that in the transverse direction of an Airy beam (cubic phase mask). The beam can be approximated by an annular mask multiplied by a 1D Gaussian function, so the focal plane is a circular Bessel function convolved with a 1D Gaussian. The two limiting cases are that of a Bessel beam, and that of a cosine beam, i.e. a fringe pattern.

Ojeda-Castañeda *et al.* proposed using a radially-symmetric phase filter that produces an axial amplitude identical to the transverse variation of the cubic phase masks of Dowsky and Cathey [228,Dowsky, 1995 #2811].

The Pearcey integral is the 1D Fourier transform of the complex exponential phase function for spherical aberration (a fourth order phase variation) and defocus, i.e. a 1D beam with spherical aberration [229]. Janssen considered the Pearcey-type integral given by the Hankel transform of a similar phase variation, and Kirk *et al.* computed and plotted this integral [230, 231].

Ring *et al.* introduced and investigated the Pearcey beam and the Pearcey-Gauss beam[232].

## 16 Photonic nanojets

Since photonic nanojets were first described by Chen et al. in 2004 [233], their paper has received more than a thousand citations according to Google Scholar, demonstrating that this phenomenon has caught the imagination of researchers in the field. A photonic nanojet is a narrow beam of light produced at the back surface of a dielectric sphere when it is illuminated with a plane wave. Potential applications include microscopy, lithography, coupling to optical fibers, materials processing, and so on. However, many of the subsequent citations have not mentioned how the photonic nanojet effect fits in with classical optics, whereas often claims are made about superfocusing or superresolution, which obviously should require comparison with the classical case. Here, we review a few concepts that predate the introduction of photonic nanojets, with a view to placing the technology in its historical setting. The topics we discuss are the relationship between finite difference time domain (FDTD) modeling and electromagnetic theory, Bessel beams and Bessel-gauss beams, the aplanatic points of a dielectric sphere, the Luneburg lens, and ball lenses.

The first paper on photonic nanojets by Chen et al. presented a FDTD treatment of focusing of a plane wave by a dielectric cylinder [233]. FDTD is a powerful modeling technique for investigating electromagnetic fields in nanophonics and microphotonics. A characteristic of the time domain nature of FDTD is that specification of the excitation field is separated temporally from the resultant scattered field. However, it should be borne in mind that it is an electromagnetic method, based on Maxwell's equations, and therefore cannot produce results that are in disagreement with the predictions of basic electromagnetism. Important features of FDTD models are the prediction of evanescent waves or plasmonic interactions, but evanescent waves are also well known, of course, in classical optics.

If a dielectric sphere is illuminated by a plane wave, the field on the far side of the sphere includes evanescent components, but these decay quickly over a distance smaller than the wavelength. At a distance greater than that for the evanescent components to have decayed, the field consists of only propagating components, and therefore exactly the same field as exhibited by the photonic nanojet could have been synthesized using macroscope optical components, such as a lens with an appropriate pupil modulation. So a photonic nanojet should not be considered as focusing better than is possible by classical means (i.e. as achieving superfocusing), although it might be a more convenient way of producing such a beam. Several papers have discussed the properties of the intermediate- or mid-field, which occurs between the near-field, where there are evanescent components, and the far-field, where there are no evanescent components. The question arises as to whether the intermediate-field contains evanescent components or not. Of course, the answer to this question depends on what one means by the term intermediate-field. Actually, we did discuss this point for the particular case of a sub-wavelength hole in a conductor [234].

A lens with a circular focuses light to a small spot, known in the paraxial approximation as an Airy disk, with a cylindrical intensity variation given by

$$I(\rho) = \left[\frac{2J_1(k\rho\sin\alpha)}{k\rho\sin\alpha}\right]^2, \tag{55}$$

where $J_1$ is a Bessel function of order 1, $k = 2\pi/\lambda$, $\rho$ is the cylindrical radius, and $\alpha$ is the semi-angular aperture of the lens. As the numerical aperture is increased, the spot becomes smaller, and eventually is modified slightly in shape as a result of nonparaxial (i.e $\sin\theta \approx \theta$ is not valid) and polarization effects. For the paraxial case, the uniform (unweighted) circular pupil produces the maximum possible intensity at the focal point (the Luneburg apodization condition). We take the full-width at half maximum (FWHM) of the central lobe in this scalar case for $\alpha = 90°$ as the classical limit for focusing.

If the pupil has a central circular obstruction, although the intensity at the focus for a given input power is lower, the FWHM of the central lobe of the point spread function (PSF) becomes narrower. The limiting case is when we are left with just a narrow ring, and the intensity point spread function is

$$I(\rho) = J_0(k\rho\sin\alpha) \tag{56}$$

This expression was first described by Rayleigh, but Airy had previously published a similar result (in series form), although he did not mention Bessel functions as these were given this name much later. We take the FWHM for $\alpha = 90°$ as the classical Bessel limit for focusing, a case which has been discussed by Zhu *et al.* [235, 236].

It was shown many years ago that the FWHM of the central lobe can be made arbitrarily narrow by appropriate weighting of the pupil, but with a strong increase in the side lobe strength [237, 238]. Hence for an unlimited field of view, and for a filter that is non-negative, it is accepted that a pupil filter cannot sharpen the PSF smaller than the Bessel beam [239].

If we turn to focusing with a cylindrical lens, the intensity line spread function is

$$I(x) = \left[\frac{\sin(kx\sin\alpha)}{kx\sin\alpha}\right]^2. \tag{57}$$

The analogy of the Bessel beam for cylindrical focusing is a fringe pattern:

$$I(x) = \cos^2(kx\sin\alpha). \tag{58}$$

This last expression gives a minimum size for features in a nonsymmetric diffraction pattern.

Traditionally, the size of the FWHM of a feature is specified as $\lambda/N$, where $N$ is a parameter. The values of $N$ for the four cases considered above are shown in Table 1.

Focusing is related to, but not the same as imaging. Note that the FWHM of the PSF is not in general the same as the Rayleigh two-point resolution criterion or the Abbe resolution limit for imaging. The FWHM for a circular aperture gives $N = 1.94$, for any degree of coherence. For two-point resolution in incoherent imaging, $N = 1/0.61 = 1.64$, while for coherent imaging the generalized Rayleigh two-point resolution gives $N = 1.22$. For the spatial frequency cut-off, $N = 2$ for incoherent imaging and $N = 1$ for coherent imaging.

The values for $N$ in Table 1 can be compared with the values in papers on photonic nanojets. One of the smallest reported is by Wu *et al.* [240], with $N = 2.58$, which is still, however, smaller than the value 2.79 for the limiting value for a scalar Bessel beam. So although photonic nanojets can be smaller in width than a classical Airy disk, they do not reach the limiting size of a Bessel beam.

**Table 1.** Values of $N$, where FWHM $= \lambda/N$.

|   | Airy disk | Bessel beam | Cylindrical | Fringe |
|---|-----------|-------------|-------------|--------|
| $N$ | 1.94 | 2.79 | 2.26 | 4 |

The Bessel beam is actually known to be not physically realizable, as it contains infinite energy. A physically realizable generalization in the paraxial regime is the Bessel-Gauss beam, which has a waist given by the product of a $J_0$ and a Gaussian [5]. By altering the relative scaling, a range of beams with different focusing properties can be generated, to fit to experimental or simulated nanojets:

$$I(\rho) = J_0^2(k\rho\sin\alpha)\exp\left[-(bk\rho\sin\alpha)^2\right]. \tag{59}$$

The peak intensity has been normalized to unity, and two parameters are incorporated, where the FWHM is primarily fixed by $\sin\alpha$ (within 1% for $0 \le b \le 0.137$), and the value of $b$ fixes the first sidelobe level. The variation in the strength of the first side lobe with $b$ is shown as the black solid curve in Fig. 12.

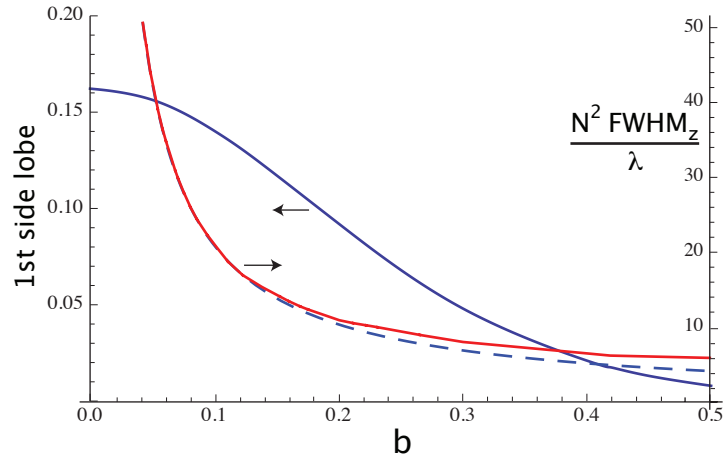

Fig. 12. The relative strength of the first side lobe as a function of the parameter (black solid curve). The variation of $N^2 FWHM_z / \lambda$ is also shown (the red solid curve shows the approximate variation, while the blue dashed curve shows the exact behaviour).

The amplitude of a Bessel-Gauss beam in the paraxial approximation can be written [5]

$$U(v,u) = \frac{e^{ikz}}{1+ib^2u} J_0\left(\frac{v}{1+ib^2u}\right) \exp\left[-\frac{b^2v^2 - iu}{2(1+ib^2u)}\right]. \tag{60}$$

where $v = k\rho\sin\alpha$ and $u = kz\sin^2\alpha$. The dimensionless variables $v, u$ are Born and Wolf's optical coordinates [30].

The intensity variation along the axis is given by [5]

$$I(u) = \frac{1}{1+b^4u^2} \exp\left[-\frac{b^2u^2}{1+b^4u^2}\right]. \tag{61}$$

The central axial lobe is found to be approximately given by (for $I < 0.2$)

$$I(u) \approx \exp(-b^2u^2), \tag{62}$$

so that the axial length of the beam is inversely proportional to $b$ and $\sin^2\alpha$. The FWHM in the axial direction is

$$FWHM_z = \frac{\sqrt{\ln 2}}{\pi b \sin^2\alpha} \lambda = \frac{0.265}{b\sin^2\alpha}\lambda. \tag{63}$$

The value of $b$ can thus alternatively be fitted for a known beam length. For small $b$,

$$\frac{N^2 FWHM_z}{\lambda} = \frac{2.06}{b}, \tag{64}$$

as shown by the blue dashed curve in Fig. 12. The red solid curve shows the exact behaviour. The lateral and axial dimensions of a focal spot are related to the first moment and second central moment, respectively, of the pupil function expressed as a function of $\rho^2$. Thus the Bessel-Gauss beam is a good approximation to any focal structure as long as it is symmetrical about the focal plane and not too flat along the axis. For more general behaviour, a beam with an additional one or two parameters, based on the third and fourth central moments, is needed to account for skewness and kurtosis.

It has been claimed that a microspherical ball lens can result in superresolution [241]. If converging light is incident on the point a distance $rn$ from the centre of a ball lens, a dielectric sphere radius $r$ and refractive index $n$, then the light is focused without aberration at the point $r/n$ from the centre of the sphere. The aplanatic points of the sphere are the basis of the design of the aplanatic front element of microscope objectives. If the radius of convergence is increased, the focused spot moves closer to the back surface of the sphere, and spherical aberration is introduced. It is well known that spherical aberration can be used to create an axicon [70], and that an axicon produces an approximation to a Bessel beam, so the fact that the focused spot for collimated illumination is like a Bessel beam should be no surprize.

## 17 Discussion

An account of the historical development of Bessel beams, and related topics such as axicons and pupil fulters, has been presented. It is hoped that this will also provide an appreciation of the principles involved, and improve understanding of the basic physics. In particular, the discussion of the connection with the spot of Arago is perhaps novel. The fact that the connection with spherical aberration appears a few times in the text is also worthy of note.